**Geological Realism in Hydrogeological and Geophysical Inverse Modeling: a Review**

Niklas Linde[1]*, Philippe Renard[2], Tapan Mukerji[3], Jef Caers[3]

[1]Applied and Environmental Geophysics Group, Institute of Earth Sciences, University of Lausanne, Switzerland;

[2]Stochastic Hydrogeology Group, Centre for Hydrogeology and Geothermics (CHYN), University of Neuchâtel, Switzerland;

[3]Stanford Center for Reservoir Forecasting, Department of Energy Resources Engineering, School of Earth Sciences, Stanford University, California.

* Corresponding author: Niklas Linde

University of Lausanne

Géopolis - bureau 3779

CH-1015 Lausanne

Email :    Niklas.Linde@unil.ch

Phone :       +41 21 692 4401

Fax :          +41 21 692 44 05

**Abstract.**

Scientific curiosity, exploration of georesources and environmental concerns are pushing the geoscientific research community towards subsurface investigations of ever-increasing complexity. This review explores various approaches to formulate and solve inverse problems in ways that effectively integrate geological concepts with geophysical and hydrogeological data. Modern geostatistical simulation algorithms can produce multiple subsurface realizations that are in agreement with conceptual geological models and statistical rock physics can be used to map these realizations into physical properties that are sensed by the geophysical or hydrogeological data. The inverse problem consists of finding one or an ensemble of such subsurface realizations that are in agreement with the data. The most general inversion frameworks are presently often computationally intractable when applied to large-scale problems and it is necessary to better understand the implications of simplifying (1) the conceptual geological model (e.g., using model compression); (2) the physical forward problem (e.g., using proxy models); and (3) the algorithm used to solve the inverse problem (e.g., Markov chain Monte Carlo or local optimization methods) to reach practical and robust solutions given today's computer resources and knowledge. We also highlight the need to not only use geophysical and hydrogeological data for parameter estimation purposes, but also to use them to falsify or corroborate alternative geological scenarios.

# 1. Introduction

Geophysical data help to understand geological processes and to test scientific hypotheses throughout the Earth Sciences, while also providing critical information and constraints for forecasting and management of subsurface formations (e.g., oil and gas reservoirs, mineral prospects, aquifers, and the critical zone). The processing of virtually all geophysical surveys involves inversion, a computational process in which measurement responses (e.g., signals in time and space for seismic and electromagnetic data) are translated into multi-dimensional images of physical properties (e.g., seismic wavespeed, density, electrical conductivity) (Menke, 1989; Tarantola, 2005) or into properties of direct relevance for geological applications (e.g., lithotype, porosity, fluid saturation) (Bosch, 1999, 2004, 2010). Subsurface heterogeneity, signal attenuation, averaging inherent to the underlying physics (e.g., diffusion), incomplete data coverage and noisy data limit the scale at which these properties can be resolved (Backus and Gilbert, 1970).

Solute transport in the subsurface can be highly sensitive to geological features (e.g., fractures (Abelin et al., 1991) or connected high conductivity forms (Zheng and Gorelick, 2003)) at scales below the resolution limits offered by geophysical sensing. Resolution-limited geophysical models alone are thus often inadequate for applications related to mass transfer in the subsurface (oil, gas, water). Even if improved geophysical acquisition systems and imaging algorithms allow resolving ever-finer details, fundamental resolution limits persist. At the high resolution necessary for flow- and transport modeling, the geophysical inverse problem has a possibly infinite set of solutions.

This non-uniqueness is traditionally overcome by using an optimization approach with a model regularization term, thereby focusing solely on model features that are necessary to explain the geophysical data (Constable et al., 1987). Such a regularization term generally lacks geological justification and results in blurry models that are overly smooth and geologically unrealistic (Ellis and Oldenburg, 1994). One step forward is to artificially introduce fine-scale information by adapting multi-Gaussian geostatistical models that describe the correlation between two points in space throughout the volume of investigation (we refer to Chilès and Delfiner (2012) for a general introduction to geostatistics). However, similar to the overly smooth models obtained by regularized inversion, the multi-Gaussian framework is often insufficient to describe realistic geological structure and especially those impacting flow responses (Gómez-Hernández and Wen, 1998; Journel and Zhang, 2006; Feyen and Caers, 2006; Kerrou et al., 2008).

In many cases, the measured hydrogeological or geophysical data can be complemented by ancillary information on the heterogeneity of subsurface formations that is obtained from borehole data, analog outcrops or databases of previously studied sites. Expert knowledge is also important. For example, sedimentologists may provide geological descriptions of the architecture of rock facies, their mutual spatial relationships, geometrical constraints or rules of deposition. In applications where supporting data are sparse and the geological context is unclear, it is perhaps even more important to assimilate and formally test competing conceptual geological models (Feyen and Caers, 2006; Park et al, 2013; Linde, 2014).

This review describes existing approaches to incorporate prior geological understanding in the inversion of geophysical and hydrogeological data to better predict subsurface flow- and transport processes at relevant temporal and spatial scales. This assimilation problem is at the forefront of many exploration, environmental, and research challenges of relevance for the Earth Sciences. Research in the area is very active, but publications are widely spread over various discipline journals with little interaction across disciplines (e.g., oil/gas vs. groundwater). Only a few attempts have been made to bridge these community gaps (e.g., Hyndman et al., 2007).

The presentation is structured as follows. Section 2 formulates the inverse problem as the integration of the information offered by geophysical and hydrogeological data, their relationship, and an underlying conceptual Earth model. Section 3 describes approaches to create geologically realistic priors and how to generate geologically realistic realizations by sampling this prior. Section 4 introduces approaches on how to parameterize models and propose model updates that are representative samples of a geologically realistic prior. Section 5 reviews how the inverse problem can be solved in the general case using sampling techniques and under more approximate conditions using stochastic search and optimization. Section 6 proposes two alternative strategies for bringing the various pieces (section 3-5) together in solving practical field cases. Section 7 provides concluding remarks.

## 2. The inverse problem

### *2.1. General formulation*

Tarantola and Valette (1982) formulated the general nonlinear inverse problem as a combination of the information provided by *N* data, **d**, by a priori information about *M* model parameters, **m**, and by theories that relate the two $p(\mathbf{m},\mathbf{d})$. In the following, a slightly less general formulation is considered that is based on a traditional Bayesian framework (Jaynes and Bretthorst, 2003).

The posterior probability density function (pdf) $p(\mathbf{m}|\mathbf{d})$ is

$$p(\mathbf{m}|\mathbf{d}) = \frac{p(\mathbf{d}|\mathbf{m})p(\mathbf{m})}{p(\mathbf{d})}, \qquad (1)$$

where $L(\mathbf{m}|\mathbf{d}) \equiv p(\mathbf{d}|\mathbf{m})$ is the likelihood function that typically summarizes the statistical properties of the error residuals between observed and simulated data and $p(\mathbf{m})$ is the prior pdf. The evidence $p(\mathbf{d})$ is important for model selection and averaging, but it can be neglected when considering a fixed model parameterization. In this case, the unnormalized density suffices

$$p(\mathbf{m}|\mathbf{d}) \propto L(\mathbf{m}|\mathbf{d})p(\mathbf{m}). \qquad (2)$$

The solution to the inverse problem can be represented as a closed-form expression of $p(\mathbf{m}|\mathbf{d})$, an approximation based on samples from this distribution or one representative model obtained by optimization.

### *2.2. The likelihood*

The forward problem consists of simulating the data response $\mathbf{d}^{sim}$ of a proposed model $\mathbf{m}^{prop}$

$$\mathbf{d}^{\text{sim}} = g(\mathbf{m}^{\text{prop}}). \tag{3}$$

The forward simulator $g(-)$ typically involves numerical simulations based on a physical theory (e.g., the advection-dispersion equation to predict tracer breakthrough curves or the electromagnetic wave equation to simulate ground-penetrating radar responses).

Assuming that measurement and modeling errors follow a Gaussian distribution, the likelihood function is

$$L(\mathbf{m}|\mathbf{d}) = \frac{1}{(2\pi)^{N/2} \det(\mathbf{C}_D)^{1/2}} \exp\left(-\frac{1}{2}\big(g(\mathbf{m}) - \mathbf{d} - \mathbf{b}_D\big)^T \mathbf{C}_D^{-1} \big(g(\mathbf{m}) - \mathbf{d} - \mathbf{b}_D\big)\right), \tag{4}$$

where $\mathbf{C}_D$ is a covariance matrix given by the sum of the covariance matrices describes modeling $\mathbf{C}_T$ and observational errors $\mathbf{C}_d$ (e.g., Tarantola, 2005) and $\mathbf{b}_D = \mathbf{b}_T + \mathbf{b}_d$ describing bias terms associated with modeling and observational error distributions that are not centered on zero (Hansen et al., 2014).

It is common practice to assume that both data and modeling errors are uncorrelated, thus, making $\mathbf{C}_D$ a diagonal matrix. This choice is often made out of convenience and because it is challenging to determine proper error models of field data (observational and geometrical errors) and forward solvers (simplified physics, numerical approximations, effects of parameterization, etc.). Gaussian error models are very sensitive to outliers and alternative distributions, for example, symmetric exponentials may provide more robust results (e.g., Claerbout and Muir, 1973). Furthermore, replacing $\mathbf{C}_D$ with a diagonal matrix and ignoring bias terms can lead to important inversion artifacts (Hansen et al., 2014), but determining $\mathbf{C}_D$ and $\mathbf{b}_D$ can be very challenging in practice. One approach is to use a computationally expensive, but physically correct forward simulator, to build an error model that is used in subsequent inversions that rely on simplified forward models (Hansen et al., 2014). Another approach is to approximate these errors with an assumed functional form, while inferring parameter values (e.g., those in an autoregressive model) during the inversion process (Schoups and Vrugt, 2010).

Furthermore, statistical rock physics models can be included in the likelihood function (e.g., Doyen et al., 1989) to link physical properties (sensed by geophysical data) and hydrogeological target properties. These relationships are often more straightforward when dealing with time-lapse data (i.e., monitoring of geophysical variables over time). Statistical rock physics is an area of active research. At present, the spatial support and correlation of the scatter in rock physics relationships, their scaling as a function of observational scale, and how parameters vary in space are often largely unknown.

*2.3. The prior*

In its simplest form, the *M* model parameters refer to material properties in a regular mesh. In this case, the standard multi-Gaussian description of the prior pdf $p(\mathbf{m})$ takes a similar form as the likelihood function (Tarantola, 2005)

$$p(\mathbf{m}) = \frac{1}{(2\pi)^{M/2} \det(\mathbf{C}_\mathrm{M})^{1/2}} \exp\left(-\frac{1}{2}(\mathbf{m}-\mathbf{m}_0)^T \mathbf{C}_\mathrm{M}^{-1}(\mathbf{m}-\mathbf{m}_0)\right), \quad (5)$$

with $\mathbf{C}_\mathrm{M}$ the model covariance matrix describing the spatial correlation between model cells and $\mathbf{m}_0$ the expected value of the model parameters. Assuming a multi-variate distribution of the prior will strongly influence the spatial characteristics of the posterior solutions. A Gaussian prior with a Gaussian likelihood function leads, in the linear case, to an explicit pdf for the posterior which is also Gaussian (e.g., Tarantola, 2005). Similarly, a Gaussian mixture prior with a Gaussian mixture likelihood gives an explicit expression for the posterior Gaussian mixture pdf (Grana et al., 2012a). However, complex priors describing realistic geological settings are often poorly described by an explicit pdf and alternatives are needed.

## 3. Simulation of geologically-based prior model realizations

Hydrogeological subsurface heterogeneity has traditionally been mostly modeled using multi-Gaussian spatial laws (e.g., Kitanidis, 1997). Such a representation has many advantages, including mathematical tractability and parsimony, since the spatial dependency between points within a model is completely defined by its mean and covariance function, which can be directly estimated from subsurface data. In the 1990's, it became clear that this framework was insufficient to adequately cover all possible heterogeneity patterns found in geological formations (Gómez-Hernández and Wen, 1998; Journel and Deutsch, 1993; Zinn and Harvey, 2003). Inadequate heterogeneity models may lead to systematic bias in model predictions and underestimation of uncertainties, especially when large data sets are available (Kerrou et al., 2008; Scheibe and Chien, 2003). This section focuses on different approaches to integrate geological understanding in prior model realizations.

### *3.1. Geologically realistic heterogeneity models*

In the framework of this paper, a subsurface model is considered geologically realistic when it explicitly integrates geological understanding (expertise, outcrops, databases) in the form of rules, patterns and geometries using quantitative methods of varying complexity. For example, a mere interpolation of geological facies is not considered a plausible geological model if it does not include some general information about the facies architecture derived from geological reasoning.

An important, but largely unanswered, question is how to define quantitative criteria to detect if the geometry and structures of a given heterogeneity model are plausible? Expert knowledge can be used to reject models that are too simplistic or do not include features that are characteristic of a certain environment. For example, the geomorphology of channels can be described by a set of morphometric indicators (Howard and Hemberger, 1991). The comparison of such indicators derived from proposed models and field observations could allow distinguishing those that are more realistic than others. While this is a promising approach, general quantitative indicators and corresponding databases of relevant indicators are still needed to provide objective criteria (e.g., Keller, 1992; Jung, and Aigner, 2012).

### 3.2. Process-based modeling

Process-based modeling consists of simulating the geological processes that lead to geological formations and the resulting internal heterogeneity is obtained as a by-product of these processes (Paola, 2000). Certain process-based simulators solve a set of partial differential equations that describe sediment transport, compaction, diagenesis, erosion, dissolution, etc. (Gabrovsek and Dreybrodt, 2010; Koltermann and Gorelick, 1992; Nicholas et al., 2013), others use cellular automata (Figure 1). Process-based simulations allow for analyzing processes that are difficult or impossible to observe at the appropriate time and spatial scales through physical experiments (Gabrovsek and Dreybrodt, 2010). They also allow reconstructions of geological patterns from the paleo-history of sedimentary basins (Goncalves et al., 2004). Their main limitations are that the data required to constrain boundary conditions and source terms for a given site are often not available and long computing times limit their usefulness for stochastic simulations and inversion. These techniques are also poorly suited for conditioning to direct and indirect data and therefore they are not described in more details here. Nevertheless, process-based models are the most advanced tools available today to produce geologically realistic models.

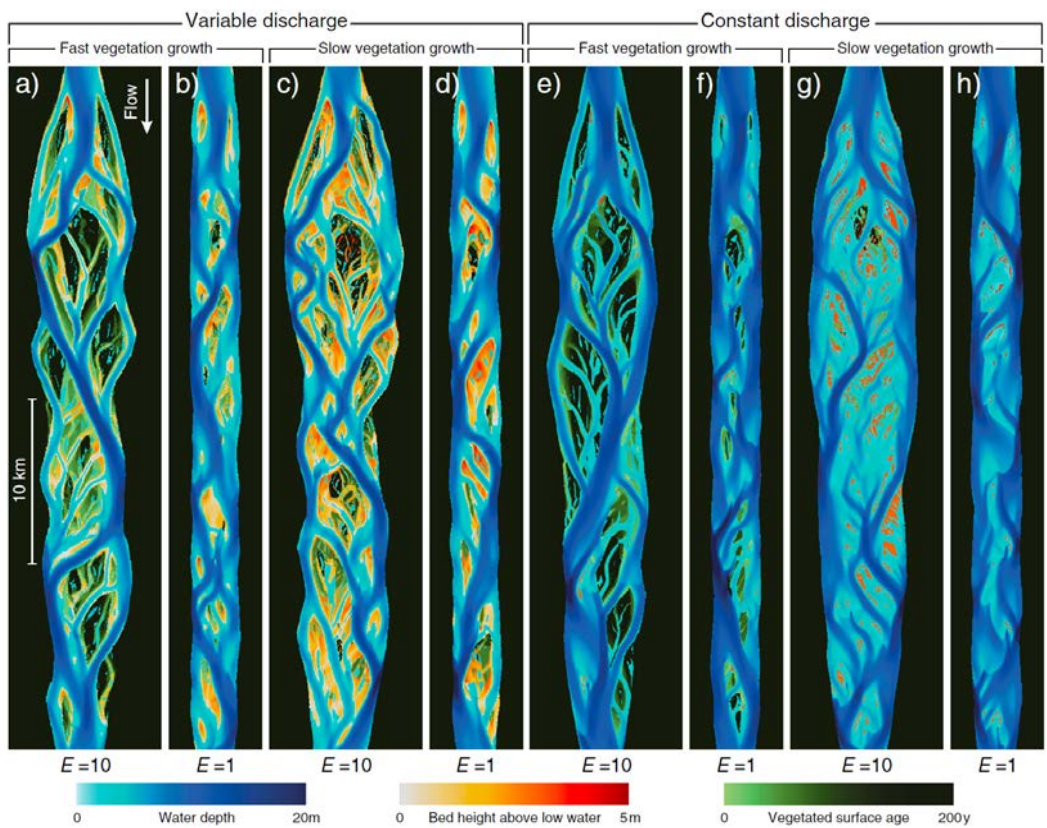

**Fig. 1.** Process-based simulations based on cellular automata that describes the topography of a braided system as a function of various controlling factors, such as bank erodability ($E$), vegetation growth, and discharge (Nicholas et al., 2013).

This type of process-based models is very useful to answer scientific questions about the factors that control geological evolution. Unfortunately, it is most difficult to condition process-based simulations to site-specific data since there is no direct deterministic link between the algorithmic variables and the resulting simulated field.

### 3.3. Object and pseudo-genetic models

Object and pseudo-genetic methods provide structure-imitating realizations (Koltermann and Gorelick, 1996) and offer a compromise between numerical efficiency and geological realism. A wide range of methods has been proposed for different types of geological environments (Deutsch and Wang, 1996; Jussel et al., 1994; Nordahl and Ringrose, 2008; Ramanathan et al., 2010; Scheibe and Freyberg, 1995; Webb and Anderson, 1996). They usually decompose the heterogeneity into a set of individual structures (architectural elements) with sizes, positions and orientations drawn from statistical distributions. The simplest techniques consider only one element type (e.g., a sinusoidal sand channel, a fracture, or a clay lens) that the algorithm places in space according to prescribed rules. Modern algorithms may include many architectural elements (e.g., channels, levees, crevasse splays, clay drapes), the shapes of the objects are more flexible and the relations between architectural elements are accounted for, as are their temporal evolution. For instance, a fracture model may include fracture growth and interactions that mimic mechanical processes (Davy et al., 2013). Similarly, as shown in Figure 2, the processes of channel evolution through time (e.g., sedimentation, avulsion) can be accounted for while simulating the objects (Lopez et al., 2008; Pyrcz et al., 2008). This leads to geological simulations that are not only fast, but also realizations that display a similar degree of geological richness as those obtained by time-consuming process-based models. Such ideas have also been used to develop 3D models of karst networks (Borghi et al., 2012; Rongier et al., 2014) by accounting for pre-existing geology, fracturing, and phases of karstification without solving the flow, transport, and calcite dissolution equations (see Figure 3). This type of approach results in conduit geometries that are highly realistic and that are expected to better describe connectivity and groundwater flow than those obtained based on purely statistical arguments.

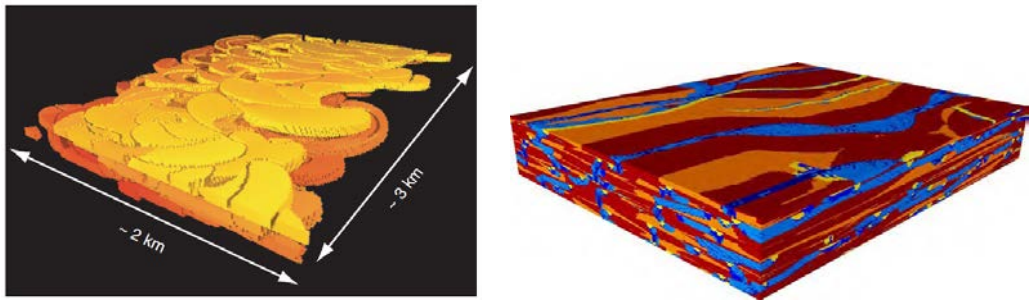

**Fig. 2.** Object based simulation of a fluvial system. Left: Evolution of a meandering river system after a period of 10'000 years (Lopez et al., 2008). Right: A 3D distribution of the facies resulting from the same model (Data courtesy: Mines Paris Tech).

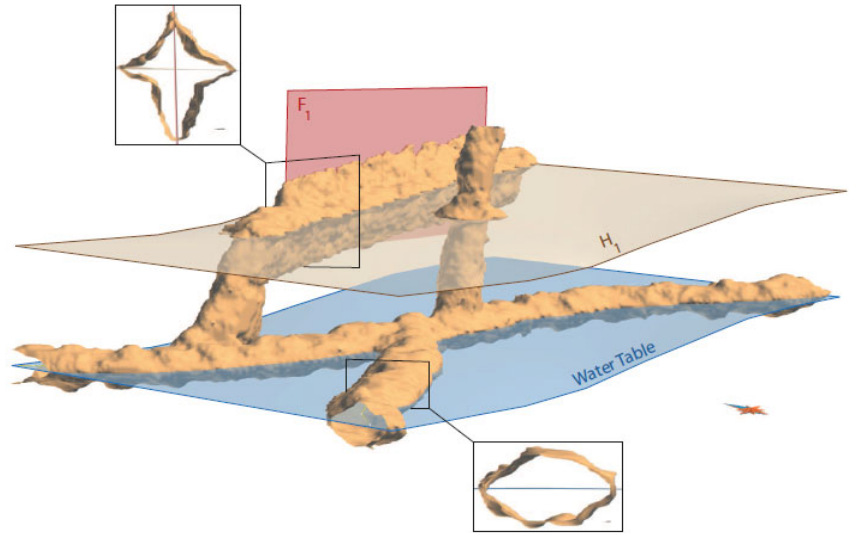

**Fig. 3.** Example of the simulation of the geometry of a karstic cave using a mixture of an object based technique and genetic concepts (Rongier et al., 2014).

A current trend to simulate simple structures, such as the centerline of a channel or the topography of a depositional surface, is to use (within the pseudo-genetic method) more advanced spatial statistics. For example, training-image based methods (described in section 3.4) can be used in combination with process-based techniques to train a multiple-point statistics (MPS) algorithm to model lobes that are stacked to create a deltaic structure (Michael et al., 2010) or a braided river system (Figure 4). Images of real channels can be used to train MPS methods to simulate realistic channels within an object based simulation approach (Mariethoz et al., 2014).

Object-based methods can result in highly realistic descriptions of subsurface heterogeneity. They are fast and can be conditioned (sometimes with difficulty) to local measurements. One of the main issues with this approach is that models are often specific for one type of geological environment only and a large number of parameters need to be determined from analog sites, thereby emphasizing the need for databases (Eschard et al., 2002; Jung and Aigner, 2012).

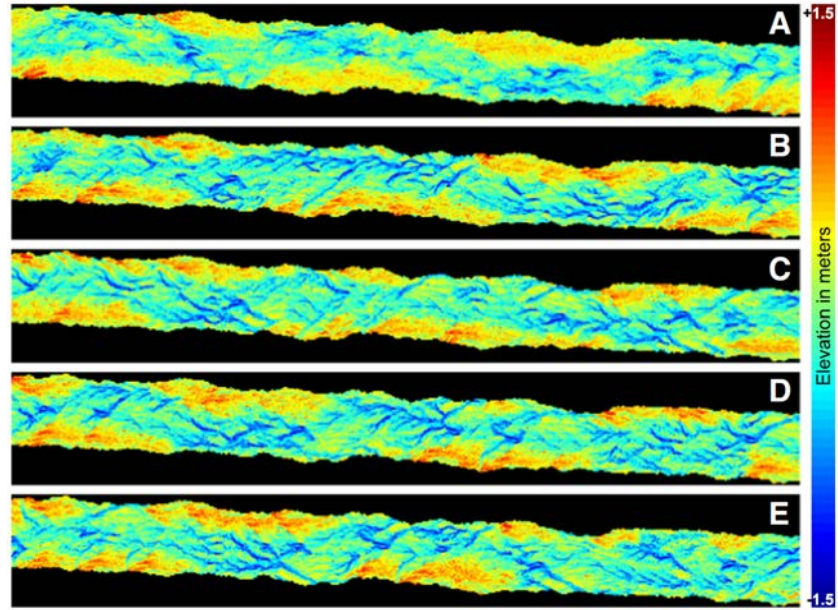

**Fig. 4.** Example of the simulation of successive topographies by using the direct sampling MPS approach, at each time step, to model the internal geological heterogeneity of braided river systems (Pirot et al., 2014).

*3.4. Training image based models*

A new class of structure imitating approaches emerged 20 years ago (Guardiano and Srivastava, 1993; Mariethoz and Caers, 2015). It uses a training image that represents a fully informed description of how the subsurface may look like, but with the locations of different repeating structures being unknown. The concept of a training image can be seen as a vehicle to convey the prior conceptual geological knowledge (Journel and Zhang, 2006) that is to be combined with other sources of information (e.g., boreholes, outcrop, etc.) via the simulation algorithm (Caers et al., 2006). The first successful simulation algorithm (snesim) based on these ideas works with high order conditional statistics or multiple-point statistics (MPS) derived from the training image (Strebelle, 2002). The training image is analyzed and the number of occurrences of each pattern is stored in a search tree. A pattern is defined as a group of cells with certain values and a certain geometrical relation. During the simulation, the search tree is used to estimate conditional probabilities by retrieving all patterns that are compatible with the available data (Strebelle, 2002). The SNESIM algorithm is restricted to categorical images with a few categories because of computational (memory) limitations. Several alternative and improved methods have been proposed (see review by Hu and Chugunova, 2008). The concept of a training image opened up a whole set of possible simulation methods. Indeed, why not use techniques derived from pattern recognition, texture synthesis and machine learning algorithms (Mariethoz and Lefebvre, 2014)?

It is now possible to apply training image based techniques with both continuous and categorical variables (Arpat and Caers, 2007; Tahmasebi et al., 2012a; Zhang et al., 2006). For example, the direct sampling algorithm allows simulations within a multivariate framework with both categorical and continuous variables (Mariethoz et al., 2010a; Meerschman et al., 2014).

In the last ten years, the focus has been on making algorithms more efficient and better at reproducing patterns in the training image (Strebelle and Cavelius, 2014). Parallel and graphics processing unit (GPU) versions of various algorithms have been implemented (e.g. Huang et al., 2013; Peredo et al., 2014; Straubhaar et al., 2011, 2013; Tahmasebi et al., 2012b). New approaches derived from image analysis and pattern simulations are currently explored (Mahmud et al., 2014) (Figure 5).

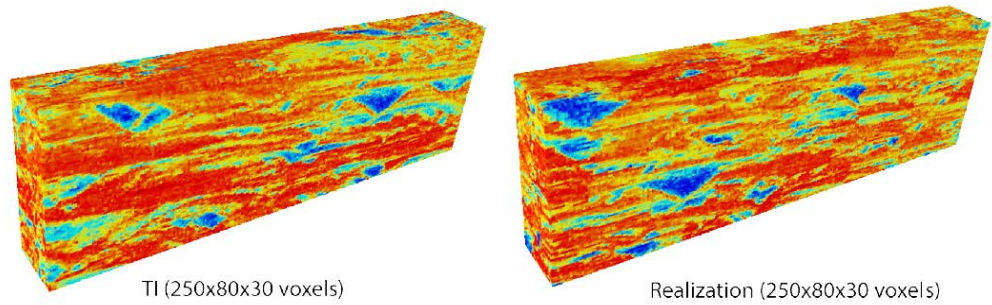

**Fig. 5.** Training image based simulation of a continuous variable representing grain sizes using the image quilting approach. Left: Training image obtained from flume experiment; Right: one simulation (Mahmud et al., 2014).

Training image based techniques are general and can be applied, in theory, to all kinds of geological environments. It is also straightforward to account for local conditioning data. The only requirement is a suitable training image, but obtaining this image can be challenging (Boucher, 2009; Chugunova and Hu, 2008; Comunian et al., 2012, 2014; Straubhaar et al., 2011). Common approaches include using a process-based or an object-based method (see sections 3.2 and 3.3; Comunian et al., 2014; Maharaja, 2008), outcrop data (Huysmans and Dassargues, 2009; Kessler et al., 2013) or a pre-existing geological model (Blouin et al., 2013; Hu et al., 2014). It should also be acknowledged that certain continuous and elongated geological structures are still difficult to model with current state-of-the-art MPS methods.

Markov random fields (MRF), originally developed in statistical physics and image processing, have been used to define geologic prior models with spatially correlated categorical variables such as different lithologies. For a spatial process, the Markovian property results in the full conditional distributions being specified by the conditional distribution given only the values in a spatial neighborhood, often described on a grid or a lattice. MRF models are based on rigorous mathematical and probability theory foundations and can account for multiple point statistics. In this sense MRFs are parametric multiple-point statistical models. MRF models of moderate size can be sampled from with exact forward-backward algorithms but for large grids Markov chain Monte Carlo sampling has to be used. This works in 2D but Markov chain Monte Carlo (MCMC) sampling becomes impractical in most 3D applications, as the convergence can be very slow. Two modifications of the general MRF model have been aused in reservoir modeling. In the so-called profile Markov models (e.g. Ulvmoen and Omre, 2010), the depth dimension is separated from the lateral directions. The depth dimension is sampled by a direct solver like a forward-backward algorithm while the MCMC sampler iterates over the 2D lateral dimensions. The convergence is much faster than a full 3D MCMC sampling. Another subclass of MRF models that has seen some 3D applications are the so-called Markov mesh models (MMM) that make MRFs more applicable by introducing an ordering on the grid. Markov mesh models are a type of partially ordered Markov models, which consider the conditional distribution for a cell given the cells with a lower order and not the entire general neighborhood as in a MRF model. Stien and Kolbjornsen (2011) show applications for MMM for facies modeling (see also Chapter 4 in Mariethoz and Caers, 2015). Although MRF models are based on a solid theoretical foundation and only require a few parameters, it is challenging to construct MRF models that produce geologically realistic realizations. For instance, a large neighborhood is required to reproduce channel structures seen in reservoirs. With large neighborhoods, careful approximations and intensive computing is required and MCMC algorithms tend to be very slow. Inferring the parameters from sparse data is also problematic and often the parameters are inferred from a training image. Directly using the training image in multiple-point geostatistical algorithms offers a better practical alternative.

### *3.5. Variogram based models*

Variogram-based approaches are widely used, but they are often insufficient to capture the complexity of geological structures. Sequential indicator simulations (SIS) (Gómez-Hernández and Srivastava, 1990) or transition probability based techniques, such as T-Progs (Carle and Fogg, 1996), were remarkable advances in the 1990's and they are still among the most popular techniques to model geological heterogeneity (dell' Arciprete et al., 2012; Falivene et al., 2006; Lee et al., 2007; Refsgaard et al., 2014). Unfortunately, they cannot properly reproduce curvilinear features, such as channels (Strebelle, 2002) or more complex structures and they do not include conceptual geological information beyond simple constraints on the dimension and relations between structures. They are also limited in simulating realistic subsurface hydraulic connectivity, which often has considerable impact on fluid flow.

A method of increasing popularity is the truncated pluri-Gaussian approach (Armstrong et al., 2003; Emery, 2007; Le Loc'h and Galli, 1997; see Fig. 6). Its principle is to model two (or more) multi-Gaussian fields with underlying variograms. These fields are then transformed into a single categorical field using truncation rules. The truncation rules offer a means to describe possible relations between geological facies. For example, in a fluvial environment it is possible to impose channels to be surrounded by levees, which in turn are surrounded by a flood plain (e.g., Mariethoz et al., 2009). As compared to SIS or T-Progs, the inference of the underlying variogram is more complex since the multi-Gaussian fields are not observed and an iterative method must be employed. The advantage of the method is that it requires only a very general geological concept. It can handle strong non-stationarity along the vertical and horizontal directions and is capable of generating complex patterns. An issue is that the contact relations defined in the truncation rule are isotropic. For fluvial systems, this implies that levees are usually found all around the channels, including at the top of the channels, which is geologically unrealistic.

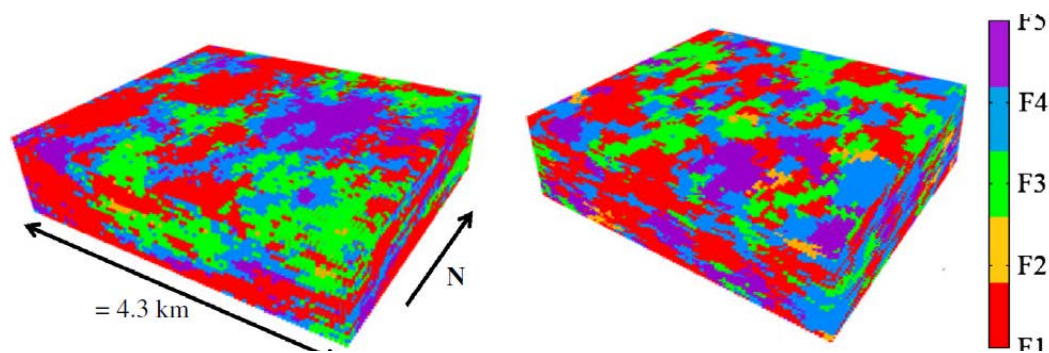

**Fig. 6.** Visual comparison of (a) a truncated pluri-Gaussian simulation and (b) a sequential indicator simulation (Perulero Serrano et al., 2014). The relations (contacts) between the facies are less erratic on the truncated pluri-Gaussian simulation than on the sequential indicator simulation. For example, the blue facies is usually located in between the green or red and violet on the left image, while all types of contact can be observed on the right picture. Nevertheless, the overall geological realism is low.

## 4. Model parameterization and perturbation

### 4.1. *Introduction*

The previous section described strategies to produce geologically realistic subsurface models. The present section focuses on how to parameterize inverse problems and how to perturb model realizations to ultimately derive subsurface models that are geologically realistic and in agreement with site-specific data. For the simplest geological models, only a few parameters are sufficient, such as the thickness of a geological layer, the dip of a fault, or the diameter of a sand lens. In these cases, solving the inverse problem is straightforward by basically perturbing model parameters iteratively (e.g. Wellmann et al., 2014). However, in most cases the degree of complexity of the geological models is much higher (see section 3) and other techniques are required. In particular, geological models may typically include millions of discretized elements that need to be populated by parameter values. Reducing the number of unknowns in the inverse problem is often necessary to avoid prohibitive computing times, but it can be challenging to preserve geological realism when working with a low dimensional representation.

### *4.2 Model reduction*

Model reduction aims at finding a model vector $\mathbf{m}_{red}$ with $M_{red} \ll M$ that provides an equivalent representation (or at least describes the most salient features of the subsurface heterogeneity) of the model vector $\mathbf{m}$ that describes the subsurface at the highest possible resolution (e.g., Figure 5). In general terms, model reduction consists in defining a mapping function $f(\cdot)$ between $\mathbf{m}$ and a lower-dimensional $\mathbf{m}_{red}$

$$\mathbf{m}_{\text{red}} = f(\mathbf{m}). \tag{6}$$

Model reduction is traditionally achieved by defining regions of constant properties or relying on the framework of multi-Gaussian fields using various schemes (see review by de Marsily et al., 1999) such as pilot points (RamaRao et al., 1995), self-sequential calibration (Gómez-Hernández et al., 1997), or gradual deformation (Hu, 2000).

Another approach consists of analyzing a set of geological models (see section 3) using image compression techniques. Examples include the use of wavelets (Sahni and Horne, 2005), Karhunen-Loeve or Discrete Cosine Transforms (Jafarpour and Mclaughlin, 2007), or Singular Value Decomposition (SVD and K-SVD) (Khaninezhad et al., 2012). The main advantages of those transforms are their generality and numerical efficiency. In these cases, the mapping is usually based on a linear combination of base vectors that are gathered in a matrix $\mathbf{F}$:

$$\mathbf{m}_{\text{red}} = \mathbf{Fm}. \tag{7}$$

A possible extension of this framework is kernel based Principle Component Analysis (KPCA; Sarma et al., 2008) which introduces non-linearity and therefore more flexibility in the transform, but render the back-transformation (termed pre-image problem) more complex. In all cases, the back transform usually produces a continuous field and not a discrete map of facies. Therefore, the examples available so far in the literature are restricted to rather simple geological models (e.g. Khaninezhad and Jafarpour, 2014; Sarma et al., 2008).

### *4.3 Prior-based model perturbation*

It is essential to consider the spatial characteristics of the geologically realistic models when making model proposals $\mathbf{m}_{prop}$. A perturbation or transition technique is thus needed that allows moving, during the inversion process, from a current model $\mathbf{m}_{pres}$ to $\mathbf{m}_{prop}$ while preserving most of the model structure:

$$\mathbf{m}_{prop} = f(\mathbf{m}_{pres}). \tag{8}$$

The function $f(-)$ is a perturbation mechanism that is not necessarily formulated analytically.

We first illustrate this principle using the multi-Gaussian case. In the gradual deformation method, a proposed model $\mathbf{m}_{prop}$ is obtained using the linear combination

$$\mathbf{m}_{prop} = \mathbf{m}_{pres}\cos(\theta) + \mathbf{m}_{random}\sin(\theta), \tag{9}$$

where $\mathbf{m}_{random}$ is a random realization of a multi-Gaussian field and $\theta$ is an angle. This weighted sum ensures that the proposed model $\mathbf{m}_{prop}$ belongs to the Gaussian prior. The difference compared with $\mathbf{m}_{pres}$ grows with $\theta$ and $\mathbf{m}_{prop}$ becomes independent of $\mathbf{m}_{pres}$ when $\theta = 90°$. It is straightforward to extend the gradual deformation to truncated Gaussian and pluri-Gaussian fields (Hu et al., 2001). The same principle can also be generalized to combine the uniform random numbers that are underlying most stochastic techniques, for example, to deform object-based simulations of fractures (Jenni et al., 2007).

The probability perturbation method (PPM) is similar to the gradual deformation method in principle but offers a different perspective (Caers, 2007; Caers and Hoffman, 2006). Instead of combining simulations directly or modifying the underlying random numbers, PPM takes a linear combination of two probability fields to obtain a single probability field that is then used as soft data to guide MPS simulations. This model perturbation technique is rather general and applicable for object based, pluri-Gaussian, and MPS models. Consider the case with a single global perturbation parameter *r* with **m** a binary model. To achieve a perturbation, the current realization is perturbed using a model of probabilities **p** defined on the same grid as **m**:

$$\mathbf{p}(r,\mathbf{m}) = (1-r)\mathbf{m}_{pres} + r\, p_m, \tag{10}$$

where $p_m$ is the marginal distribution, in this case simply the global proportion. This probability model is then used as soft probability to generate a new realization. The value of *r* regulates the degree of model perturbation from one model realization to another (Figure 7). To allow for more flexibility in the perturbation, regions, each with a different *r*, can be introduced. This achieves a regional perturbation where some regions may change more than others. Grana et al., (2012b) use the PPM method to generate facies realizations conditioned to seismic data, but the geologic prior is described simplistically by a variogram based model.

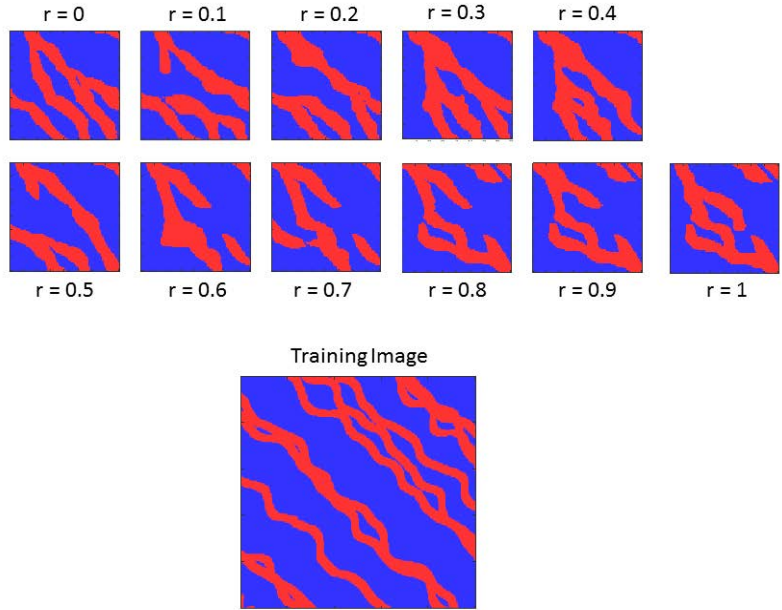

**Fig. 7.** Example of perturbations from one MPS realization drawn from the prior into another prior sample by means of the probability perturbation method (adapted from Mariethoz and Caers, 2015).

It is not always possible that the perturbation mechanism (eq. 8) can be defined analytically. It is then necessary to obtain $\mathbf{m}_{prop}$ by resimulating a fraction of the model cells in $\mathbf{m}_{pres}$ conditional to those that are left unchanged. The simplest transition between two models is simply to re-simulate one random model cell at a time, but this is very slow as many transitions steps are needed to create a significant model perturbation. To accelerate the transition, the blocking moving window method re-simulates a whole portion of the model domain at each iteration (Fig. 8), while the remaining part of the model domain and all field observations are kept as conditioning data. The location, and possibly the dimension of the window (usually a rectangle), is changed randomly. This approach was pioneered by Fu and Gómez-Hernández (2009) for multi-Gaussian fields, before being applied to MPS simulations (Alcolea and Renard, 2010; Hansen et al., 2012, 2013).

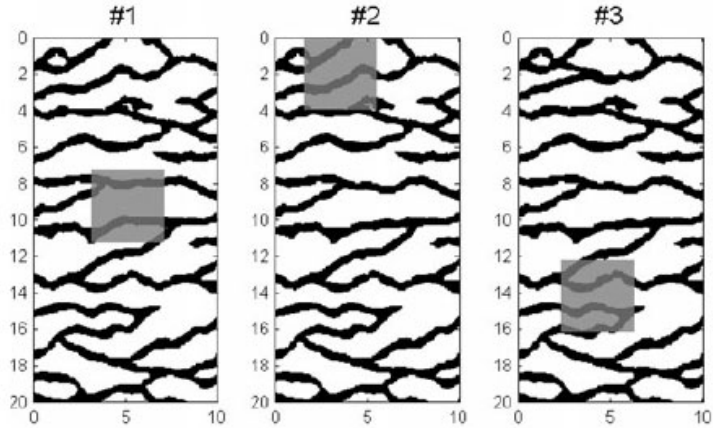

**Fig. 8.** The blocking moving window method uses sequential resimulation of a part of the domain (Hansen et al., 2012).

In contrast to the blocking window method, the Iterative Spatial Resampling (ISR) method (Mariethoz et al., 2010b) consists of re-simulating grid cells throughout the model domain. At each step, a fixed number of random points are selected and used as conditioning data for the next iteration. The perturbation is large when there are few conditioning points and small when there are many conditioning points. The location of the conditioning points can be completely random or focused in regions were the model is expected to be reliable (Jeong et al., 2012).

**5. Inversion with complex priors**

The inverse problem can be solved under different limiting assumptions. The formulation can be rigorous and hence often slow, or simplistic and possibly inaccurate. It is important to use a formulation that is adapted to the problem at hand and the available computing resources. This can be addressed by considering algorithms that allow varying the trade-off between competing elements of the inverse problem. For a given computational budget, it is often revealing to analyze the results obtained by varying the error description, the accuracy of the forward model, the rock physics relationships, the training image and the inverse algorithm itself.

Below, we briefly introduce three basic categories of approaches to solve inverse problems: sampling based methods, stochastic search methods and optimization methods.

### *5.1. Sampling based methods*

Sampling based methods aim at approximating $p(\mathbf{m}|\mathbf{d})$ by drawing random samples from this distribution. This is either done by evaluating random and independent samples from the prior or by importance sampling that preferentially sample significant areas of $p(\mathbf{m}|\mathbf{d})$.

Rejection sampling is the only exact sampler, but it is often computationally infeasible. It proceeds by repeating the following two steps

1. Draw a sample $\mathbf{m}_{\text{prop}}$ from $p(\mathbf{m})$

2. Accept this proposal as a draw from $p(\mathbf{m}|\mathbf{d})$ with probability

$$p = \min\left\{1, \frac{L(\mathbf{m}_{\text{prop}}|\mathbf{d})}{S_L}\right\}, \tag{11}$$

where $S_L$ is the supremum of the likelihood function, which is generally unknown and must therefore be set to a large value. In a recent case study, Dorn et al. (2013) used rejection sampling to obtain a set of discrete fracture network models that agreed with both hydrogeological and geophysical data. The rejection sampler works well if the prior model space is small, but is unfeasible for most practical applications. Rejection sampling is primarily used as a benchmark sampler to evaluate the performance of other sampling methods.

Markov chain Monte Carlo (MCMC) methods, such as Metropolis sampling (Hastings, 1970) or Gibbs sampling (Geman and Geman, 1984) can be used to sample from the posterior distribution. A chain of model realizations are generated, in which $\mathbf{m}_{prop}$ is dependent on the previous model $\mathbf{m}_{pres}$. A very simple way to inject such dependency in spatial models is to retain some part of the previous realization as conditioning data (see section 4.3).

Mosegaard and Tarantola (1995) developed an extended version of the Metropolis algorithm that is applicable to large spatial problems with complex priors, whose analytical form is not available. In particular their method can be applied to MPS-based priors. It proceeds by repeating the following three steps

1. Re-simulate parts of $\mathbf{m}_{pres}$ using any algorithm that produce geologically realistic models (e.g., an MPS algorithm) to obtain $\mathbf{m}_{prop}$ (see section 4.3)
2. Accept $\mathbf{m}_{prop}$ with probability

$$p = \min\left\{1, \frac{L(\mathbf{m}_{prop}|\mathbf{d})}{L(\mathbf{m}_{pres}|\mathbf{d})}\right\} \qquad (12)$$

3. If accepted then $\mathbf{m}_{pres} \leftarrow \mathbf{m}_{prop}$

Conditional simulations ensure that $\mathbf{m}_{prop}$ is drawn from $p(\mathbf{m})$. It is possible to optimize the acceptance rate by determining the appropriate size of the blocking moving window or the fraction of model cells to be updated in the ISR method (section 4.3). Figure 9 shows an application of the Metropolis sampler with iterative spatial resampling. The data are the seismic responses modeled using an approximate 2-D Born filter on the MPS realizations where seismic velocities differ considerably between facies (channel/background). The results of the Metropolis sampler compare well with the rejection sampler.

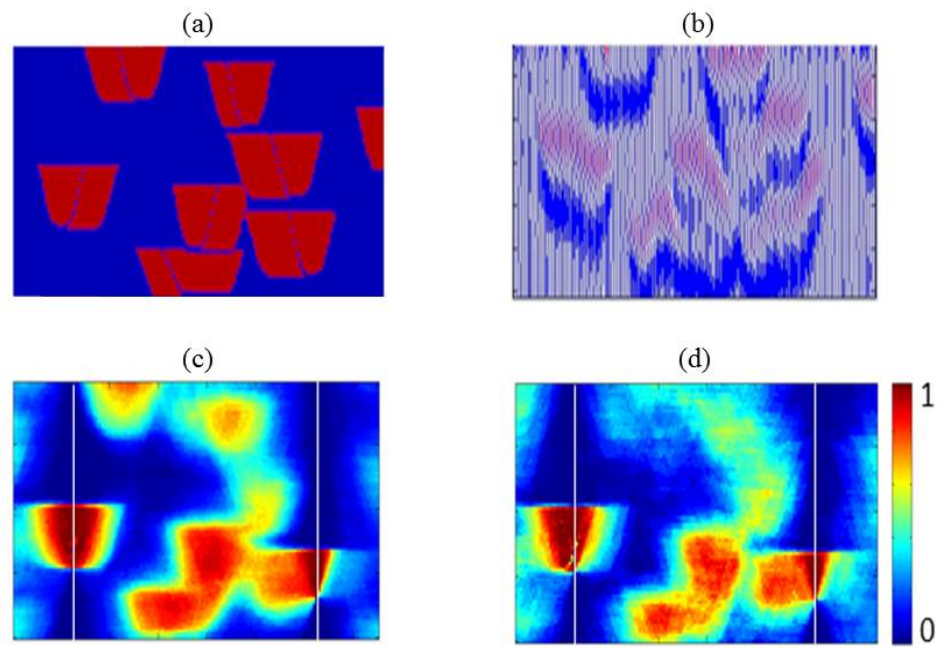

**Fig. 9.** Comparison of Metropolis sampler (using iterative spatial resampling) with rejection sampling. (a) Reference Facies model (b) corresponding synthetic amplitude data (c) Ensemble average of the rejection sampler and (d) ensemble average of the Metropolis sampler. Shown are the two well locations (white lines) where facies conditioning data are present. The rejection sampler required about 100,000 forward model evaluations while the Metropolis sampler took only 500 evaluations (from Jeong et al., 2012). The prior model consists of a realization generated from a training image (not shown).

Ulvmoen and Omre (2010) used Markov random fields to define a prior model for lithologies, and then used block Gibbs simulation to sample from the posterior, conditioned to seismic amplitude versus offset (AVO) data. In a hierarchical Bayesian setting, Rimstad and Omre (2010) also relied on Markov random fields for the prior and used MCMC to sample from the posterior distribution of facies and rock properties conditioned to seismic data. The likelihood function consists of rock physics relations and a linearized convolution model for the seismic data.

One problem with the extended Metropolis algorithm is that the acceptance rate of proposals often becomes very low, and the algorithm becomes inefficient, when using many model parameters and large data sets. This can be avoided by using only small model updates, but this increases the risk of the chain getting stuck in local minima. More efficient MCMC algorithms exist that use local gradients in the likelihood function (stochastic Newton; e.g., Martin et al., 2012), several chains with different temperatures (parallel tempering; e.g., Sambridge, 2014), the history of sampled points (adaptive MCMC; e.g., Haario et al., 2001) or the state of parallel chains to make efficient model proposals (differential evolution; e.g., Vrugt et al., 2009), etc. However, adapting these methods in the context of geologically realistic priors is non-trivial. Recently, Lochbühler et al. (2015) used an ensemble of training images to define a reduced model parameterization and a prior distribution in terms of summary statistics. They combined adaptive MCMC, differential evolution and subspace sampling as implemented in the DREAM$_{(ZS)}$ algorithm (Laloy and Vrugt, 2012) to invert for the porosity field and a rock physics transform using crosshole geophysical data.

### *5.2. Stochastic search methods*

One of the problems with sampling methods is the CPU time. MCMC methods are very slow, and they become impractical for applications where the forward model takes a few hours to run. Simplifications are thus often needed to address realistic three-dimensional case-studies, but it is generally unclear which are the assumptions that can be made and those that should be avoided. Intuition built from low-dimensional examples is often not applicable in high parameter dimensions.

Stochastic search methods are designed to find the global minimum of an objective function and they have been used widely for geophysical applications. These methods include simulated annealing techniques (SA), genetic algorithms (GA), the neighborhood search algorithm (NA) and particle swarm optimization (PSO). Sen and Stoffa (2013) give an overview of global optimisation methods and their applications in geophysical inversion. Stochastic search based methods are faster than sampling methods, but only sample approximately from the posterior pdf. They provide multiple realizations that:

- are samples from the spatial geological prior as specified with any of the methods in section 3;
- the data are matched up to a specified error level defined through an objective function.

Often, the major subjectivity in inversion lies in the choice of the prior. When the model formulation itself is highly subjective one may question the need for rigorous sampling and it might become interesting to instead explore multiple realizations obtained by stochastic search methods. An example of this approach is given by Gonzalez et al. (2008) who used a training image and multiple-point simulations to create samples from the prior. These are then matched to seismic data within a specified error level, giving multiple realizations that honor the spatial geologic prior. Their algorithm uses a sequential trace-by-trace approach, with rock physics relations and convolutional seismic forward modeling. Below, we provide further examples of stochastic search methods in presence of a geologically realistic prior.

The Neighborhood algorithm (NA, Sambridge, 1999a,b) was originally proposed for seismological inverse problems and later applied to flow inverse problems (Christie et al., 2006). The method derives an approximate misfit surface using previously evaluated misfit functions, and, based on the approximate misfit, identifies multiple parameter combinations that are likely to achieve a minimum misfit. The approximate misfit surface is constructed by compartmentalizing the model space into Voronoi cells. The neighborhood algorithm was originally designed for parameterized problems, where parameters could, for example, be object shape parameters, or layer parameters, all listed as a vector. In case of geological prior formulated with training images, Suzuki and Caers (2008) extended NA to priors involving MPS training images by parameterizing the problem using distances between model realizations. NA only requires the definition of a distance to move about and search for models sampled from the prior with low misfit. Suzuki and Caers (2008) used this approach to solve a multiphase subsurface flow inverse problem with a set of 81 training images representing alternative prior geological scenarios.

The ensemble Kalman filter (EnKf; Aanonsen et al., 2009; Evensen, 1994; Houtekamer and Mitchell, 1998; Oliver and Chen, 2011) is an approach to data-based forecasting that has recently gained attention for subsurface inverse problems. EnKf is a recursive filter operation where a mismatch in the data is used to adjust the model by a linear update operation. In its most basic formulation, EnKf assumes a multi-Gaussian distribution on model and data variables and a linear relationship between all variables. Several authors have studied these limitations and proposed extensions of the EnKf in cases when the prior geological uncertainty can no longer be realistically modeled with a multi-Gaussian distribution. Sarma and Chen (2009) proposed a machine learning approach whereby the EnKf is applied after a kernel transformation, possibly including a Karhoenen-Loeve expansion (Sarma et al., 2008). The problem with this approach lies in the back-transformation (to the actual model space). This back-transformation is non-unique, hence constitutes an inverse problem on its own, subject to the same geological constraints as the initial problem. Jafarpour and Khodabakhshi (2011) applied the EnKf to soft (auxiliary) variables that controls the generation of geostatistical realizations (in their case MPS models). Hu et al. (2013) proposed to apply the EnKf to the uniform random numbers used to generate the geostatistical realizations by means of a gradual deformation-based parameterization. Other approaches rely on transforming the non-Gaussian local distributions into Gaussian ones on which then the EnKf can be applied (Zhou et al., 2011). Alternatively, Zhou et al. (2012) proposed a pattern-based search method in combination with direct sampling (Mariethoz et al., 2010a) to directly generate an ensemble of realizations that match the data and reflect geological patterns. In that spirit, they only retain the idea of using an ensemble but do not rely on linear updates or transformation of space. While the linear, Gaussian form provides (multi-Gaussian) posterior uncertainty that is well understood, the extended techniques only offer partial and approximate posterior uncertainties.

*5.3. Optimization methods*

In some practical cases, it may be useful to obtain just one solution, for example, the solution **m** that corresponds to the maximum of $p(\mathbf{m}|\mathbf{d})$ or a maximum a-posterior (MAP) solution. Lange et al. (2012) defined a MAP solution for TI-based priors in a way that is reminiscent of methods of regularization for solving inverse problems (Tikhonov and Arsenin, 1977): a data mismatch term is coupled with a regularization term with the aim to induce some desired property onto the solution:

$$\mathbf{m}^{\text{MAP}} = \underset{\mathbf{m}}{\text{argmin}} \left( \frac{1}{2} \| g(\mathbf{m}) - \mathbf{d}_{\text{obs}} \|_{\mathbf{C}_D}^2 + \alpha\, h(\mathbf{m}) \right), \tag{13}$$

where the subscript indicates that the data are weighted with respect to $\mathbf{C}_D$. In traditional regularization, $h(\mathbf{m})$ is often used to induce smoothness by a discretized gradient operator and the weight $\alpha$ is used to regulate the degree of smoothness. In Lange et al. (2012), the formulation is generalized to consider regularization with respect to the training image. Instead of stating a degree of smoothness (such as by means of a covariance or derivatives), the MAP solution is enforced to have patterns similar to the training image. The function $h(\mathbf{m})$ is defined to be a measure of dissimilarity between the training image patterns and the patterns of any realization generated from it. To obtain a summary of pattern frequencies, one can use a multi-point histogram, which consists of a simple counting of the occurrence of pixel-configurations within a given template or neighborhood. A squared difference in counts would then form such a function $h(\mathbf{m})$. Perturbation methods such as simulated annealing can be used to find the MAP solution. Figure 10 shows clearly the influence of the regularization, where a MAP solution with insufficient regularization deviate from the training image patterns and provide unrealistic looking solutions.

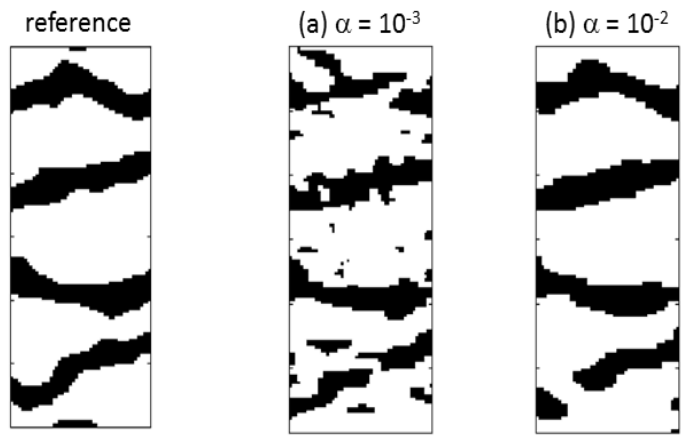

**Fig. 10:** (a) A reference model and (b, c) MAP solutions with different weights given to patterns found in a training image. With a sufficient weight, it is possible to obtain (c) models that are consistent with the training image patterns (from Lange et al., 2012).

## 6. Workflows

### 6.1 Introduction

The inversion of data and the creation of subsurface models that "match" the data is rarely an end goal. Practical field applications start with a purpose or decision question: what are the subsurface hydrogeological models ultimately used for and what decisions are required? Such a question of purpose leads to a number of possible follow-up questions. What subsurface structures impact this decision the most? What are the most useful data to answer a scientific question or characterize a target of interest? For example, should the focus be on static geophysical data (i.e., acquisition at one given time) that mainly provide information about the geological structure, or time-lapse data that are better suited to inform about dynamic processes? What is the best combination of geophysical and traditional hydrogeological data (i.e., pumping tests, tracer tests)? What is the relative value of emerging techniques that provide extensive high-resolution data on hydrogeological properties (e.g., Bohling et al., 2012) compared with upscaled and more integrated data (e.g., geophysics, tracer and pumping tests)? What is the relative information content of the prior with respect to the data? If the prior is uncertain (this is always the case), is it better that the final solutions are potentially more affected by the prior than tens of thousands of seismic traces or extensive hydraulic tomography tests? Or, should the immense richness of seismic data dominate over geological knowledge? These questions can only be appropriately addressed by considering flexible workflows.

Alternative workflows for the joint inversion of several data sources have recently been proposed by several authors both in the groundwater and petroleum literature (e.g. Bosch et al., 2010; Castro et al., 2009; Doyen, 2007; Ferré et al., 2009; Herckenrath et al., 2013; Hinnell et al., 2010). These authors cover the methodological aspect of joint use, whether through sequential or joint assimilation of data sources. Most workflows do not cover or include geological information sources and uncertainty assessment is primarily on the geophysical and rock physics aspects of the inverse problem. Below, we describe two alternative ways of thinking that capture most ideas in the combined literature that we elaborate upon in the next section:

- A top-down, Popper-Bayes workflow: the modeling focus lies on the hydrogeological/geological prior model and how data, both geophysical and hydrogeological, can be used to falsify geological concepts and reduce an initially large prior uncertainty.
- A bottom-up, data-focused workflow: the focus lies on data inversion, whether sequential or joint, and prior geological models are used either as constraints to the inversion or to downscale inversion results into high-resolution geological/hydrogeologic models for forecasting.

### *6.2. Top-down, Popper-Bayes*

Tarantola (2006) proposed to combine Popper's philosophy of falsification with Bayesian information theory in a geophysical context. In such an approach, the focus lies on rejecting scenarios that are incompatible with the data rather than on constructing models that match the data as well as possible (see also Oreskes et al. (1994) and Linde (2014)). To make this practical, it is necessary to construct a very wide prior of geological models that include a large set of possibilities. Such methodologies are mostly lacking in the inversion literature because it calls on a very different way of doing scientific analysis and synthesis: geological interpretation of all available data such as, for example, depositional genesis in the case of sedimentary environments (e.g., Hermans et al., 2015).

Figure 11 provides a broad overview of a strategy to achieve uncertainty and risk quantification in three stages based on the Popper-Bayes' concept. As such, it is not a formal methodology, but outlines possible combinations of methodologies depending on specific field challenges.

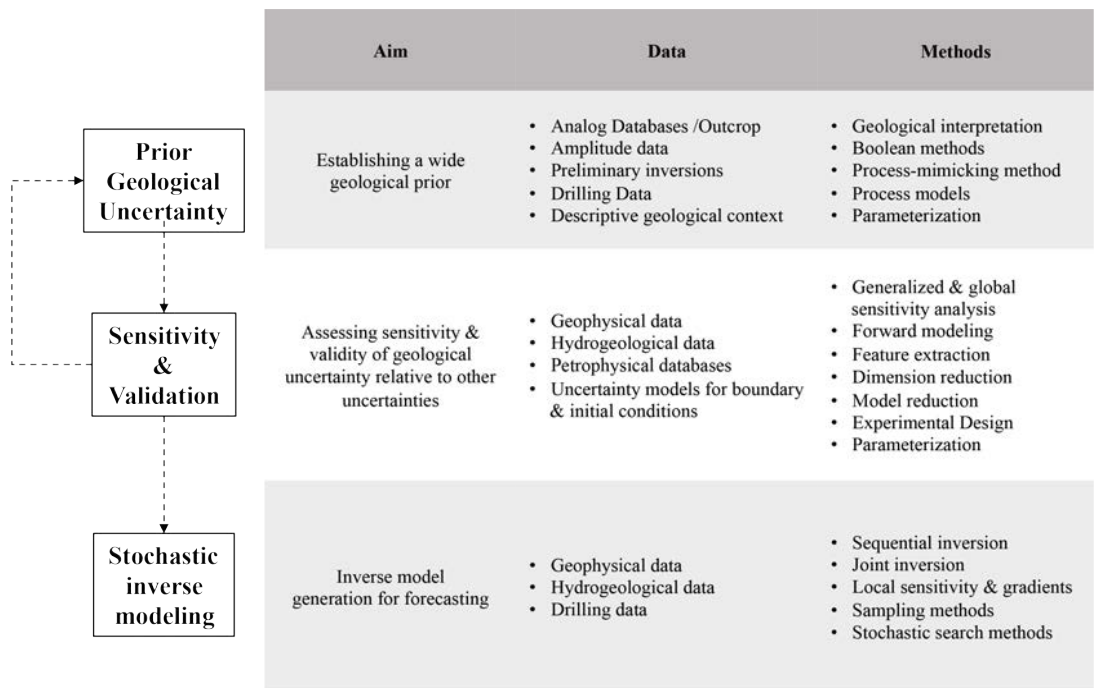

**Fig. 11:** Scheme for a general workflow on stochastic inversion and uncertainty analysis.

*Geological prior model construction.* Even in the construction of a wide geologically realistic prior, data enter the equation, in particular data obtained from drilling, logging and from geophysics. Geological reasoning does not happen in a vacuum and data should be used at this stage, but such usage is only for interpretation, not for inversion. For example, geophysical amplitude data can be used as it may reveal interpretable morphologies and shapes, although only at the resolution of that data or within a limited extent of the domain. From this data and the geological context, geologists collect information leading to statements about the possible nature of the depositional environments (alluvial, fluvial, deltaic, carbonate mound, etc.). None of these various alternatives should be eliminated a priori if data or context does not clearly support this. Within each system, variations exist due to sub-classifications within that system (see Jung and Aigner, 2012) or variations in certain parametric descriptions of the system. Analog databases, for example, on proportions, paleo-direction, morphologies and architecture of facies bodies or geological rules of association (Eschard et al., 2002; Gibling, 2006) for various geological environments (FAKT: Colombera et al., 2012; CarbDB: Jung and Aigner, 2012; WODAD: Kenter and Harris, 2006; Paleoreefs: Kiessling and Flügel, 2002; Pyrcz et al., 2008) can be consulted to define prior distributions about geometries and spatial distribution of geobodies. From a modeling perspective, one may consider two levels of uncertainty: the uncertain style or classification, often termed "geological scenario" (Martinius and Næss, 2005) and variations within that style. Boolean simulation methods and codes (Deutsch and Tran, 2002; Deutsch and Wang, 1996; Holden et al., 1998; Maharaja, 2008) as well as process-mimicking codes (e.g. Bertoncello et al., 2013; Pyrcz et al., 2009) can then be used either as prior model parameterization or for generating training images based on a selected style and on a set of sample parameters. For visualization and quality check unconditional realizations can be generated, in a hierarchical fashion: depositional style/geological scenario → geological parameters → 3D model realizations.

The rock physics properties can then be defined within layers or geobody types. The rock physics properties are often modeled using more traditional semi-variogram based methods, hence the parameterization at that level is in the form of univariate statistical parameters (histogram) and semi-variogram parameters (variogram type, ranges, anisotropy ratios). As for the geological scenarios, the prior parameters of those statistical distributions can be derived from existing databases (e.g., Comunian and Renard, 2009).

*Sensitivity and Corroboration.* In the second stage the important question of "what matters?" (sensitivity) and "is it possible?" (corroboration & rejection) are tackled. At this level, the focus is not on matching data. In sensitivity analysis, one will need to account for other types of uncertainties, not just depositional/geological ones. These uncertainties include: structural geological (if faults exist, see Caers, 2011, Chapter 8 for a review), rock physics relationship uncertainty (Mavko et al., 2009) and dynamic flow uncertainty (including boundary conditions, initial conditions, fluid properties, relative permeability, etc.). Sensitivity may be calculated with respect to data responses or to forecast responses/decision variables and often requires forward modeling. Note that a joint sensitivity of all parameters is required as many parameters across the various disciplines involved (geology, geophysics and flow) may interact with each other (Fenwick et al., 2014). Sensitivity may allow for significant model reduction as insensitive model parameters can be assigned constant values. A second geological model reduction may occur through corroboration. In the Popper-Bayes philosophy, uncertainty in the geological model is stated as independently as possible from subsurface data. As a consequence, the relationship between this wide prior and the data needs to be assessed to identify inconsistent geological assumptions (invalid scenarios) and/or to discover the most likely scenarios. Such screening does not require inversion, instead it focuses mostly on the scenario level of geological uncertainty. As such, it could rely on extracting features and global patterns from the geophysical data (e.g., wavelet coefficient histograms) and comparing them with the same features extracted from forward modeling responses from selected prior model realizations. In that context, a conditional probability can be estimated for each scenario from a few forward models based on differences between simulated responses and actual data (Park et al., 2013; Scheidt et al., 2015). Note that in practice (as indicated in Figure 11), one may need to iterate between geological prior model construction and sensitivity/corroboration, since it may occur that the stated prior cannot explain the data (prior model vs data inconsistency).

*Inversion.* The broad geological prior initially stated may now have been reduced through sensitivity analysis & corroboration to a reduced prior model, with possibly prior probabilities associated to certain geological scenarios. At this stage one also has gained more insights into the model-data-forecast relationships and thus confidence that inversion methods (section 5) will be successful in matching the data.

We argue that a Popper-Bayes approach is often highly relevant for catchment or reservoir scale problems. This workflow is also suitable for cases where non-geological modeling elements, such as rock physics uncertainty, boundary and initial condition uncertainty, chemical/physical uncertainty of fluids are prevalent. In such cases, constraints can be added to the inversion to ensure geological plausibility of the inverted model.

### 6.3 Bottom-up, data-focused workflow

The previous workflow is particularly relevant when subsurface geological heterogeneity plays an important role and the geophysical and hydrogeological data are not constraining enough to accurately image the subsurface. In other cases, there is considerable confidence on the nature of geological scenarios and only minor (parameter) variations within one single scenario are needed to describe uncertainty. Often this is true when considering practical problems that occur over smaller domains such as contaminated sites or the surroundings of a well. The bottom-up workflow can be separated into sequential and joint inversions (Bosch et al., 2010).

In sequential (hydro)geophysical inversion, geophysical data are first inverted to provide physical properties; then, rock physics is used to convert (deterministically or stochastically) the inverted physical properties into geophysical scale reservoir/hydrogeological properties which may be further downscaled using geostatistical techniques (e.g., Rubin et al., 1992; Copty et al., 1993; Cassiani and Medina, 1997). The sequential approach has been used widely for hydrocarbon reservoir modeling to integrate geophysical, well log and core data. This is often the most straightforward approach to use geophysical data for hydrogeological purposes and it may provide useful results. The main criticisms of the sequential approach are (1) that the resolution limitations of the geophysical models are typically ignored in the rock physics mapping and this may lead to unphysical results, such as, loss of mass (Day-Lewis et al., 2007); (Moysey et al., (2005) attempted to correct the rock physics mapping by accounting for resolution limitations by numerical simulations); (2) it is difficult to constrain the geophysical inversion to hydrogeological constraints (Ferré et al., 2009) and (3) the estimates are biased if the rock physics relationship is non-linear (Bosch, 2004).

Joint (hydro)geophysical inversion relies on first constructing geostatistical realizations of reservoir/hydrogeological properties, then converting those properties to physical properties and forward geophysical modeling for stochastic Bayesian inversion by sampling (e.g., Cassiani and Binley, 2005; Hinnell et al., 2010; Irving and Singha, 2010; Huisman et al., 2010) or optimization (e.g., Kowalsky et al., 2005; 2011) methods. To illustrate this approach, let us consider the simple case when the prior information is described in terms of the hydrogeological target properties $\mathbf{m}_{hydrogeology}$, the model realizations of these target properties can then be mapped into physical fields $\mathbf{m}_{geophysics}$ using an appropriate rock physics relationship, $p\left(\mathbf{m}_{geophysics} \middle| \mathbf{m}_{hydrogeology}\right)$. The simulated geophysical forward response of this proposed physical property field can then be compared with the observed geophysical data, $\mathbf{d}_{geophysics}$. In this situation, the posterior pdf is proportional to:

$$p(\mathbf{m}|\mathbf{d}) \propto p\left(\mathbf{m}_{hydrogeology}\right) p\left(\mathbf{m}_{geophysics} \middle| \mathbf{m}_{hydrogeology}\right) L\left(\mathbf{m}_{geophysics} \middle| \mathbf{d}_{geophysics}\right). \qquad (14)$$

It is seen that the inverse problem is essentially the same as in equation (2), except for the rock physics relationship, and all the approaches described in section 5 can be used.

## *7. Concluding remarks*

A large suite of tools is available to obtain increasingly realistic subsurface models that are conditioned to large sets of hydrogeological and geophysical data. However, their effective use in challenging field settings is still largely in development and computing limitations are often an issue when targeting realistic 3-D realizations and large data sets. To conclude this review, we want to highlight a few directions that could constitute important lines of research in the next decade. One crucial point is that it is often unclear to what extent small imperfections in our increasingly complex statistical and physical models affect predictions and uncertainty estimates. It is also important to find practical ways to use a Popper/Bayes perspective to assess which conceptual geological models that are in agreement with the available hydrogeological and geophysical data. If no models are found that agree with the data, this implies that there are aspects in the models that need to be improved. This iterative process is important to better describe geological heterogeneity, geophysical forward solvers, rock physics models and noise characteristics. To empower the data focused workflow, there is a need to find appropriate model reduction techniques that allow representing realistic geological heterogeneity in a relatively low-dimensional space. Another alternative could be to combine accurate, but computationally demanding, forward solvers with an approximate solution allowing to select efficiently the promising candidates among a set of possible geological models (Ginsbourger et al., 2013). While there is already a set of methods available to perturb geological models while preserving the geological structures, it is still very difficult to make these perturbations efficient in the sense that they decrease rapidly the data residuals. One way to improve this could be to use explicit formulas to relate the sensitivity of the forward problem to changes in the geometry of local inclusions in heterogeneous materials as recently proposed by Noetinger (2013). We emphasize that the goal of geological realism in hydrogeophysical inverse modeling is not per se to create geologically realistic earth models, but to enable more informed conclusions and decisions under uncertainty.

*Acknowledgements*

This work was partly funded by the Swiss National Science Foundation and is a contribution to the ENSEMBLE project (grant no. CRSI22_132249). We thank the members of ENSEMBLE for discussions that helped to shape this manuscript. We also would like to thank Ty Ferré, Grégoire Mariethoz and an anonymous reviewer for their constructive comments that helped to improve the quality of the final manuscript. This work is dedicated to the memory of PhD student Tobias Lochbühler who tragically lost his life in a mountaineering accident on 2014 July 19. Tobias was a truly wonderful person and a most gifted researcher.

*References*

Aanonsen, S. I., Nævdal, G., Oliver, D. S. & Reynolds, A. C. V., B., 2009. The ensemble Kalman filter in reservoir engineering: a review. SPE Journal, 14, 393–412.

Abelin, H., Birgersson, Moreno, L., Widén, H., Ågren, T., Neretnieks, I., 1991. A large-scale flow and tracer experiment in granite: 2. Results and interpretation. Water Resour. Res. 27, 3119–3135. doi:10.1029/91WR01404

Alcolea, A., Renard, P., 2010. Blocking Moving Window algorithm: Conditioning multiple-point simulations to hydrogeological data. Water Resour. Res. 46, W08511, doi:10.1029/2009WR007943.

Armstrong, M., Galli, A., Le Loc'h, G., Geffroy, F., Eschard, R., 2003. Plurigaussian simulations in geosciences. Springer, Berlin.

Arpat, G.B., Caers, J., 2007. Conditional simulation with patterns. Math. Geol. 39, 177–203. doi:10.1007/s11004-006-9075-3

Backus, G., Gilbert, F., 1970. Uniqueness in the inversion of inaccurate gross earth data. Philos. Trans. R. Soc. Lond. Ser. Math. Phys. Sci. 266, 123–192.

Bertoncello, A., Sun, T., Li, H., Mariethoz, G., Caers, J., 2013. Conditioning surface-based geological models to well and thickness data. Math. Geosci. 45, 873–893. doi:10.1007/s11004-013-9455-4

Blouin, M., Martel, R., Gloaguen, E., 2013. Accounting for aquifer heterogeneity from geological data to management tools. Ground Water 51, 421–431. doi:10.1111/j.1745-6584.2012.00982.x

Bohling, G.C., Liu, G., Knobbe, S.J., Reboulet, E.C., Hyndman, D.W., Dietrich, P., Butler, J.J., 2012. Geostatistical analysis of centimeter-scale hydraulic conductivity variations at the MADE site. Water Resources Research, 48, W02525. doi: 10.1029/2011WR010791.

Borghi, A., Renard, P., Jenni, S., 2012. A pseudo-genetic stochastic model to generate karstic networks. J. Hydrol. 414, 516–529. doi:10.1016/j.jhydrol.2011.11.032

Bosch, M., 1999. Lithologic tomography: From plural geophysical data to lithology estimation. J. Geophys. Res. Solid Earth 104, 749–766. doi:10.1029/1998JB900014

Bosch, M., 2004. The optimization approach to lithological tomography: Combining seismic data and petrophysics for porosity prediction. Geophysics 69, 1272–1282. doi:10.1190/1.1801944

Bosch, M., Mukerji, T., Gonzalez, E.F., 2010. Seismic inversion for reservoir properties combining statistical rock physics and geostatistics: A review. Geophysics 75, 75A165–75A176. doi:10.1190/1.3478209

Boucher, A., 2009. Considering complex training images with search tree partitioning. Comput. Geosci. 35, 1151–1158.

Caers, J., 2007. Comparing the gradual deformation with the probability perturbation method for solving inverse problems. Math. Geol. 39, 27–52. doi:10.1007/s11004-006-9064-6

Caers, J., 2011. Modeling Uncertainty in the Earth Sciences, Édition : 1. ed. Wiley.

Caers, J., Hoffman, T., 2006. The probability perturbation method: A new look at Bayesian inverse modeling. Math. Geol. 38, 81–100.

Caers, J., Hoffman, T., Strebelle, S., Wen, X.-H., 2006. Probabilistic integration of geologic scenarios, seismic, and production data—a West Africa turbidite reservoir case study. Lead. Edge 25, 240–244. doi:10.1190/1.2184087

Carle, S.F., Fogg, G.E., 1996. Transition probability-based indicator geostatistics. Math. Geol. 28, 453–476. doi:10.1007/BF02083656

Cassiani, G., and A. Binley, 2005. Modeling unsaturated flow in a layered formation under quasi-steady state conditions using geophysical data constraints, Advances in Water Resources, 28(5), 467–477, doi:10.1016/j.advwatres.2004.12.007.

Cassiani, G., Medina, M. A., 1997, Incorporating auxiliary geophysical data into ground-water flow parameter estimation, Ground Water, 35(1), 79–91, doi:10.1111/j.1745-6584.1997.tb00063.x.

Castro, S., Caers, J., Otterlei, C., Meisingset, H., Hoye, T., Gomel, P., Zachariassen, E., 2009. Incorporating 4D seismic data into reservoir models while honoring production and geologic data: A case study. Lead. Edge 28, 1498–1505. doi:10.1190/1.3272706

Chilès, J.-P., Delfiner, P., 2012. Geostatistics: Modeling Spatial Uncertainty, 2nd Edition. Wiley, 734 pp.

Christie, M., Demyanov, V., Erbas, D., 2006, Uncertainty quantification for porous media flows. J. Comp. Phys., 217, 143-158. doi:10.1016/j.jcp.2006.01.026

Chugunova, T.L., Hu, L.Y., 2008. Multiple-point simulations constrained by continuous auxiliary data. Math. Geosci. 40, 133–146.

Claerbout, J., Muir, F., 1973. Robust modeling with erratic data. Geophysics 38, 826–844. doi:10.1190/1.1440378

Colombera, L., Felletti, F., Mountney, N.P., McCaffrey, W.D., 2012. A database approach for constraining stochastic simulations of the sedimentary heterogeneity of fluvial reservoirs. AAPG Bull. 96, 2143–2166. doi:10.1306/04211211179

Comunian, A., Jha, S.K., Giambastiani, B.M.S., Mariethoz, G., Kelly, B.F.J., 2014. Training images from process-imitating methods. Math. Geosci. 46, 241–260. doi:10.1007/s11004-013-9505-y

Comunian, A., Renard, P., 2009. Introducing wwhypda: a world-wide collaborative hydrogeological parameters database. Hydrogeol. J. 17, 481–489. doi:10.1007/s10040-008-0387-x

Comunian, A., Renard, P., Straubhaar, J., 2012. 3D multiple-point statistics simulation using 2D training images. Comput. Geosci. 40, 49–65. doi:10.1016/j.cageo.2011.07.009

Copty, N., Rubin, Y., Mavko, G. 1993. Geophysical-hydrological identification of field permeabilities through Bayesian updating, Water Resour. Res., 29(8), 2813–2825, doi:10.1029/93WR00745.

Constable, S.C., Parker, R.L., Constable, C.G., 1987. Occam's inversion: A practical algorithm for generating smooth models from electromagnetic sounding data. Geophysics 52, 289–300. doi:10.1190/1.1442303

Davy, P., Le Goc, R., Darcel, C., 2013. A model of fracture nucleation, growth and arrest, and consequences for fracture density and scaling. J. Geophys. Res.-Solid Earth 118, 1393–1407. doi:10.1002/jgrb.50120

Day-Lewis, F. D., Y. Chen, and K. Singha, 2007. Moment inference from tomograms, Geophys. Res. Lett., 34(22), L22404, doi:10.1029/2007GL031621.

De Marsily, G., Delhomme, J.-P., Delay, F., Buoro, A., 1999. 40 years of inverse problems in hydrogeology. Comptes Rendus Acad. Sci. Ser. IIA Earth Planet. Sci. 329, 73–87.

Dell' Arciprete, D., Bersezio, R., Felletti, F., Giudici, M., Comunian, A., Renard, P., 2012. Comparison of three geostatistical methods for hydrofacies simulation: a test on alluvial sediments. Hydrogeol. J. 20, 299–311.

Deutsch, C.V., Tran, T.T., 2002. FLUVSIM: a program for object-based stochastic modeling of fluvial depositional systems. Comput. Geosci. 28, 525–535.

Deutsch, C.V., Wang, L., 1996. Hierarchical object-based stochastic modeling of fluvial reservoirs. Math. Geol. 28, 857–880.

Dorn, C., Linde, N., Borgne, T.L., Bour, O., de Dreuzy, J.-R., 2013. Conditioning of stochastic 3-D fracture networks to hydrological and geophysical data. Adv. Water Resour. 62, 79–89. doi:10.1016/j.advwatres.2013.10.005

Doyen, P. M., 2007, Seismic reservoir characterization: An Earth modeling perspective, EAGE publications bv., pp 255.

Doyen, P. M., T. M. Guidish, and M. H. de Buyl, 1989, Seismic discrimination of lithology in sand/shale reservoirs: A Bayesian approach SEG Technical Program Expanded Abstracts, 719-722.

Ellis, R.G., Oldenburg, D.W., 1994. Applied geophysical inversion. Geophys. J. Int. 116, 5–11. doi:10.1111/j.1365-246X.1994.tb02122.x

Emery, X., 2007. Simulation of geological domains using the plurigaussian model: New developments and computer programs. Comput. Geosci. 33, 1189–1201.

Eschard, R., Doligez, B., Beucher, H., 2002. Using quantitative outcrop databases as a guide for geological reservoir modelling, in: Geostatistics Rio 2000. Springer, pp. 7–17.

Evensen, G., 1994. Sequential data assimilation with nonlinear quasi-geostrophic model using Monte Carlo methods to forecast error statistics. Journal of Geophysical Research, 5, 143-162.

Falivene, O., Arbués, P., Gardiner, A., Pickup, G., Muñoz, J.A., Cabrera, E., 2006. Best practice stochastic facies modeling from a channel-fill turbidite sandstone analog (the Quarry outcrop, Eocene Ainsa basin, northeast Spain). AAPG Bull. 90, 1003–1029.

Fenwick, D., Scheidt, C., Caers, J., 2014. Quantifying asymmetric parameter interactions in sensitivity analysis: Application to reservoir modeling. Math. Geosci. 46, 493–511. doi:10.1007/s11004-014-9530-5

Ferré, T., Bentley, L., Binley, A., Linde, N., Kemna, A., Singha, K., Holliger, K., Huisman, J.A., Minsley, B., 2009. Critical steps for the continuing advancement of hydrogeophysics. Eos Trans. Am. Geophys. Union 90, 200–200. doi:10.1029/2009EO230004

Feyen, L., Caers, J., 2006. Quantifying geological uncertainty for flow and transport modeling in multi-modal heterogeneous formations. Adv. Water Resour. 29, 912–929. doi:10.1016/j.advwatres.2005.08.002

Fu, J., Gómez-Hernández, J.J., 2009. A blocking Markov chain Monte Carlo method for inverse stochastic hydrogeological modeling. Math. Geosci. 41, 105–128. doi:10.1007/s11004-008-9206-0

Gabrovsek, F., Dreybrodt, W., 2010. Karstification in unconfined limestone aquifers by mixing of phreatic water with surface water from a local input: A model. J. Hydrol. 386, 130–141. doi:10.1016/j.jhydrol.2010.03.015

Geman, S., Geman, D., 1984. Stochastic relaxation, Gibbs distributions, and the Bayesian restoration of images. IEEE Trans. Pattern Anal. Mach. Intell. PAMI-6, 721–741. doi:10.1109/TPAMI.1984.4767596

Gibling, M.R., 2006. Width and thickness of fluvial channel bodies and valley fills in the geological record: A literature compilation and classification. J. Sediment. Res. 76, 731–770. doi:10.2110/jsr.2006.060

Gómez-Hernández, J.J., Srivastava, R.M., 1990. ISIM3D: An ANSI-C three-dimensional multiple indicator conditional simulation model. Comput. Geosci. 16, 395–440.

Gómez-Hernández, J.J., Sahuquillo, J.E., Capilla, J.E., Stochastic simulation of transmissivity fields conditional to both transmissivity and piezometric data—I. Theory. J. Hydrol, 203, 162-174.

Gómez-Hernández, J.J., Wen, X.-H., 1998. To be or not to be multi-Gaussian? A reflection on stochastic hydrogeology. Adv. Water Resour. 21, 47–61. doi:10.1016/S0309-1708(96)00031-0

Goncalves, J., Violette, S., Guillocheau, F., Robin, C., Pagel, M., Bruel, D., de Marsily, G., Ledoux, E., 2004. Contribution of a three-dimensional regional scale basin model to the study of the past fluid flow evolution and the present hydrology of the Paris basin, France. Basin Res. 16, 569–586. doi:10.1111/j.1365-2117.2004.00243.x

González, E. F., T. Mukerji, and G. Mavko, 2008, Seismic inversion combining rock physics and multiple-point geostatistics: Geophysics, 73, no. 1, R11–R21

Grana, D., Mukerji, T., Dovera, L., and Della Rossa, E., 2012a, Sequential simulations of mixed discrete-continuous properties: Sequential Gaussian Mixture

Simulation, in Geostatistics Oslo 2012, eds. P. Abrahamsen et al., Springer Science, Dordrecht.

Grana, D., Mukerji, T., Dvorkin, J., and Mavko, G., 2012b, Stochastic inversion of facies from seismic data based on sequential simulations and probability perturbation method, Geophysics, 77, M53-M72.

Guardiano, F., Srivastava, R.M., 1993. Multivariate geostatistics: beyond bivariate moments, in: Soares, A. (Ed.), Geostatistics Troia 1992. Kluwer Academic Publishers, Dordrecht, The Netherland, pp. 133–144.

Haario, H., Saksman, E., Tamminen, J., 2001, An adaptive Metropolis algorithm, Bernoulli, 7, 223-242.

Hansen, T., Cordua, K., Jacobsen, B., Mosegaard, K., 2014. Accounting for imperfect forward modeling in geophysical inverse problems — Exemplified for crosshole tomography. Geophysics 79, H1–H21. doi:10.1190/geo2013-0215.1

Hansen, T.M., Cordua, K.S., Looms, M.C., Mosegaard, K., 2013. SIPPI: A Matlab toolbox for sampling the solution to inverse problems with complex prior information: Part 2—Application to crosshole GPR tomography. Comput. Geosci. 52, 481–492. doi:10.1016/j.cageo.2012.10.001

Hansen, T.M., Cordua, K.S., Mosegaard, K., 2012. Inverse problems with non-trivial priors: efficient solution through sequential Gibbs sampling. Comput. Geosci. 16, 593–611. doi:10.1007/s10596-011-9271-1

Hastings, W.K., 1970. Monte Carlo sampling methods using Markov chains and their applications. Biometrika 57, 97–109. doi:10.2307/2334940

Herckenrath, D., Fiandaca, G., Auken, E., Bauer-Gottwein, P., 2013. Sequential and joint hydrogeophysical inversion using a field-scale groundwater model with ERT and TDEM data. Hydrol Earth Syst Sci 17, 4043–4060. doi:10.5194/hess-17-4043-2013

Hermans, T., Nguyen, F., Caers, J., 2015. Uncertainty in training-image based inversion of hydraulic head data constrained to ERT data: Workflow and case study. Water Resources Research, 51, 5332-5352. doi: 10.1002/2014WR016460.

Hinnell, A.C., Ferré, T.P.A., Vrugt, J.A., Huisman, J.A., Moysey, S., Rings, J., Kowalsky, M.B., 2010. Improved extraction of hydrologic information from geophysical data through coupled hydrogeophysical inversion. Water Resour. Res. 46, W00D40. doi:10.1029/2008WR007060

Holden, L., Hauge, R., Skare, Ø., Skorstad, A., 1998. Modeling of fluvial reservoirs with object models. Math. Geol. 30, 473–496. doi:10.1023/A:1021769526425

Houtekamer, P., Mitchell, H. L., 1998. Data assimilation using an ensemble Kalman filter technique. Monthly Weather Review, 126, 796-811.

Howard, A., Hemberger, A., 1991. Multivariate characterization of meandering. geomorphology 4, 161–186. doi:10.1016/0169-555X(91)90002-R

Hu, L.Y., 2000. Gradual deformation and iterative calibration of Gaussian-related stochastic models. Math. Geol. 32, 87–108.

Hu, L.Y., 2008. Extended probability perturbation method for calibrating stochastic reservoir models. Math. Geosci. 40, 875–885. doi:10.1007/s11004-008-9158-4

Hu, L.Y., Chugunova, T., 2008. Multiple-point geostatistics for modeling subsurface heterogeneity: A comprehensive review. Water Resour. Res. 44, W11413. doi:10.1029/2008WR006993

Hu, L.Y., Le Ravalec, M., Blanc, G., 2001. Gradual deformation and iterative calibration of truncated Gaussian simulations. Pet. Geosci. 7, S25–S30.

Hu, L.Y., Liu, Y., Scheepens, C., Shultz, A.W., Thompson, R.D., 2014. Multiple-point simulation with an existing reservoir model as training image. Math. Geosci. 46, 227–240. doi:10.1007/s11004-013-9488-8

Hu, L. Y., Zhao, Y., Liu, Y., Scheepens, C., Bouchard, A. 2013. Updating multipoint simulations using the ensemble Kalman filter. Computers and Geosciences, 51, 7-15.

Huang, T., Li, X., Zhang, T., Lu, D.-T., 2013. GPU-accelerated Direct Sampling method for multiple-point statistical simulation. Comput. Geosci. 57, 13–23. doi:10.1016/j.cageo.2013.03.020

Huisman, J. A., J. Rings, J. A. Vrugt, J. Sorg, and H. Vereecken, 2010. Hydraulic properties of a model dike from coupled Bayesian and multi-criteria hydrogeophysical inversion, Journal of Hydrology, 380(1–2), 62–73, doi:10.1016/j.jhydrol.2009.10.023.

Huysmans, M., Dassargues, A., 2009. Application of multiple-point geostatistics on modelling groundwater flow and transport in a cross-bedded aquifer (Belgium). Hydrogeol. J. 17, 1901–1911.

Hyndman, D.W., Day-Lewis, F.D., Singha, K., 2007. Subsurface hydrology: data integration for properties and processes. American Geophysical Union.

Irving, J., and K. Singha, 2010. Stochastic inversion of tracer test and electrical geophysical data to estimate hydraulic conductivities, *Water Resour. Res.*, *46*(11), W11514, doi:10.1029/2009WR008340.

Jafarpour, B., Khodabakshi, M. 2011. A probability conditioning method (PCM) for nonlinear flow data integration into multipoint statistical facies simulation. Mathematical Geosciences, 43, 133-164.

Jafarpour, B., Mclaughlin, D., 2007. Efficient permeability parameterization with the discrete cosine transform. Society of Petroleum Engineers. doi:10.2118/106453-MS

Jaynes, E.T., Bretthorst, G.L., 2003. Probability Theory: The Logic of Science. Cambridge University Press, Cambridge, UK; New York, NY.

Jenni, S., Hu, L.Y., Basquet, R., De Marsily, G., Bourbiaux, B., 2007. History matching of a stochastic model of field-scale fractures: methodology and case study. Oil Gas Sci. Technol.-Rev. Inst. Francais Pet. 62, 265–276.

Jeong, C., Mukerji, T., Mariethoz, G., 2012. Adaptive Spatial Resampling Applied to Seismic Inverse Modeling, in: The Ninth International Geostatistics Congress, Oslo.

Journel, A., Zhang, T., 2006. The necessity of a multiple-point prior model. Math. Geol. 38, 591–610. doi:10.1007/s11004-006-9031-2

Journel, A.G., Deutsch, C.V., 1993. Entropy and spatial disorder. Math. Geol. 25, 329–355.

Jung, A., Aigner, T., 2012. Carbonate geobodies: Hierarchical classification and database – a new workflow for 3D reservoir modelling. J. Pet. Geol. 35, 49–65. doi:10.1111/j.1747-5457.2012.00518.x

Jussel, P., Stauffer, F., Dracos, T., 1994. Transport modeling in heterogeneous aquifer: 1. Statistical description and numerical generation of gravel deposits. Water Resour. Res. 30, 1803–1817.

Keller, B., 1992. Hydrogeologie des schweitzerischen Molasse-Beckens: Aktueller Wissensstand und weiterführende Betrachtungen. Eclogae Geologicae Helvetiae. 85, 611-651.

Kenter, J.A.M., Harris, P.M., 2006. Web-based Outcrop Digital Analog Database (WODAD): Archiving carbonate platform margins, in: AAPG Int. Conf., Australia, Nov. pp. 5–8.

Kerrou, J., Renard, P., Hendricks Franssen, H.-J., Lunati, I., 2008. Issues in characterizing heterogeneity and connectivity in non-multiGaussian media. Adv. Water Resour. 31, 147–159. doi:10.1016/j.advwatres.2007.07.002

Kessler, T.C., Comunian, A., Oriani, F., Renard, P., Nilsson, B., Klint, K.E., Bjerg, P.L., 2013. Modeling fine-scale geological heterogeneity-examples of sand lenses in tills. Ground Water 51, 692–705. doi:10.1111/j.1745-6584.2012.01015.x

Khaninezhad, M.M., Jafarpour, B., 2014. Sparse Randomized Maximum Likelihood (SpRML) for subsurface flow model calibration and uncertainty quantification. Adv. Water Resour. 69, 23–37. doi:10.1016/j.advwatres.2014.02.005

Khaninezhad, M.M., Jafarpour, B., Li, L., 2012. Sparse geologic dictionaries for subsurface flow model calibration: Part I. Inversion formulation. Adv. Water Resour. 39, 106–121. doi:10.1016/j.advwatres.2011.09.002

Kiessling, W. ,Flügel, E., 2002. Paleoreefs—A Database on Phanerozoic reefs. SEPM Special Publication No 72

Kitanidis, P.K., 1997. Introduction to Geostatistics: Applications in Hydrogeology. Cambridge University Press.

Koltermann, C., Gorelick, S., 1992. Paleoclimatic signature in terrestrial flood deposits. Science 256, 1775–1782. doi:10.1126/science.256.5065.1775

Koltermann, C.E., Gorelick, S.M., 1996. Heterogeneity in sedimentary deposits: A review of structure-imitating, process-imitating, and descriptive approaches. Water Resour. Res. 32, 2617–2658. doi:10.1029/96WR00025

Kowalsky, M. B., S. Finsterle, J. Peterson, S. Hubbard, Y. Rubin, E. Majer, A. Ward, and G. Gee, 2005. Estimation of field-scale soil hydraulic and dielectric parameters through joint inversion of GPR and hydrological data, Water Resour. Res., 41(11), W11425, doi:10.1029/2005WR004237.

Kowalsky, M. B., E. Gasperikova, S. Finsterle, D. Watson, G. Baker, and S. S. Hubbard, 2011. Coupled modeling of hydrogeochemical and electrical resistivity data for exploring the impact of recharge on subsurface contamination, Water Resour. Res., 47(2), W02509, doi:10.1029/2009WR008947.

Laloy, E., Vrugt, J.A., 2012, High-dimensional posterior exploration of hydrological models using multiple-try DREAM$_{(ZS)}$ and high-performance computing. Water Resour. Res., 48, W01526.

Lange, K., Frydendall, J., Cordua, K.S., Hansen, T.M., Melnikova, Y., Mosegaard, K., 2012. A frequency matching method: Solving inverse problems by use of

geologically realistic prior information. Math. Geosci. 44, 783–803. doi:10.1007/s11004-012-9417-2

Le Loc'h, G., Galli, A., 1997. Truncated plurigaussian method: theoretical and practical points of view, in: Baafi, E.Y., Schofield, N.A. (Eds.), Geostatistics Wollongong 1996. Volume 1. Kluwer Academic Publishers, pp. 211–222.

Lee, S.Y., Carle, S.F., Fogg, G.E., 2007. Geologic heterogeneity and a comparison of two geostatistical models: Sequential Gaussian and transition probability-based geostatistical simulation. Adv. Water Resour. 30, 1914–1932.

Linde, N., 2014. Falsification and corroboration of conceptual hydrological models using geophysical data. Wiley Interdiscip. Rev. Water 1. doi:10.1002/wat2.1011.

Lochbühler, T., Vrugt, J.A., Sadegh, M., Linde, N., 2015, Summary statistics from training images as prior information in probabilistic inversion. Geophys. J. Int., 201, 157-171. doi:10.1093/gji/ggv008.

Lopez, S., Cojan, I., Rivoirard, J., Galli, A., 2008. Process-based stochastic modelling: Meandering channelized reservoirs, in: Boer, P. de, Postma, G., Zwan, K. van der, Burgess, P., Kukla, P. (Eds.), Analogue and Numerical Modelling of Sedimentary Systems: From Understanding to Prediction. Wiley-Blackwell, pp. 139–144.

Maharaja, A., 2008. TiGenerator: Object-based training image generator. Comput. Geosci. 34, 1753–1761.

Mahmud, K., Mariethoz, G., Caers, J., Tahmasebi, P., Baker, A., 2014. Simulation of Earth textures by conditional image quilting. Water Resour. Res. 50, 3088–3107. doi:10.1002/2013WR015069

Mariethoz, G., Caers, J., 2015. Multiple-point geostatistics: stochastic modeling with training images. Wiley Blackwell, 364 pp.

Mariethoz, G., Comunian, A., Irarrazaval, I., Renard, P., 2014. Analog-based meandering channel simulation. Water Resour. Res. 50, 836-854. doi:10.1002/2013WR013730.

Mariethoz, G., Lefebvre, S., 2014. Bridges between multiple-point geostatistics and texture synthesis: Review and guidelines for future research. Comput. Geosci. 66, 66–80. doi:10.1016/j.cageo.2014.01.001

Mariethoz, G., Renard, P., Caers, J., 2010b. Bayesian inverse problem and optimization with Iterative Spatial Resampling. Water Resour. Res. 46, W11530, doi:10.1029/2010WR009274.

Mariethoz, G., Renard, P., Cornaton, F., Jaquet, O., 2009. Truncated plurigaussian simulations to characterize aquifer heterogeneity. Ground Water 47, 13–24. doi:10.1111/j.1745-6584.2008.00489.x

Mariethoz, G., Renard, P., Straubhaar, J., 2010a. The Direct Sampling method to perform multiple-point geostatistical simulations. Water Resour. Res. 46, W11536. doi:10.1029/2008WR007621

Martin, J., L. C. Wilcox, C. Burstedde, and O. Ghattas (2012), A stochastic Newton MCMC Method for large-scale statistical inverse problems with application to seismic inversion, *SIAM Journal on Scientific Computing*, *34*(3), A1460–A1487, doi:10.1137/110845598.

Martinius, A.W., Næss, A., 2005. Uncertainty analysis of fluvial outcrop data for stochastic reservoir modelling. Pet. Geosci. 11, 203–214. doi:10.1144/1354-079303-615

Mavko, G., Mukerji, T., Dvorkin, J., 2009. The Rock Physics Handbook: Tools for Seismic Analysis of Porous Media, Édition : 2. ed. Cambridge University Press, Cambridge, UK ; New York.

Meerschman, E., Van Meirvenne, M., Mariethoz, G., Islam, M.M., De Smedt, P., Van de Vijver, E., Saey, T., 2014. Using bivariate multiple-point statistics and proximal soil sensor data to map fossil ice-wedge polygons. Geoderma 213, 571–577. doi:10.1016/j.geoderma.2013.01.016

Menke, W., 1989. Geophysical Data Analysis: Discrete Inverse Theory. Academic Press.

Michael, H.A., Li, H., Boucher, A., Sun, T., Caers, J., Gorelick, S.M., 2010. Combining geologic-process models and geostatistics for conditional simulation of 3-D subsurface heterogeneity. Water Resour. Res. 46, W05527. doi:10.1029/2009WR008414

Mosegaard, K., Tarantola, A., 1995. Monte Carlo sampling of solutions to inverse problems. J. Geophys. Res. Solid Earth 100, 12431–12447. doi:10.1029/94JB03097

Moysey S., K. Singha, and R. Knight, 2005, A framework for inferring field-scale rock physics relationships through numerical simulation, Geophysical Research Letters, 32, DOI 10.1029/2004GRL022152.

Nicholas, A.P., Ashworth, P.J., Sambrook Smith, G.H., Sandbach, S.D., 2013. Numerical simulation of bar and island morphodynamics in anabranching megarivers. J. Geophys. Res. Earth Surf. 118, 2019–2044. doi:10.1002/jgrf.20132.

Noetinger, B. (2013), An explicit formula for computing the sensititivty of the effective conductivity of heterogeneous composite materials to local inclusion transport properties and geometry. Multiscale Modeling & Simulation, 11, 907-924.

Nordahl, K., Ringrose, P.S., 2008. Identifying the representative elementary volume for permeability in heterolithic deposits using numerical rock models. Math. Geosci. 40, 753–771.

Oliver, D. S., Chen, Y., 2011. Recent progress on reservoir history matching: a review. Computational Geosciences, 15, 185-221.

Oreskes, N., Shrader-Frechette, K., Belitz, K., others, 1994. Verification, validation, and confirmation of numerical models in the earth sciences. Science 263, 641–646.

Paola, C., 2000. Quantitative models of sedimentary basin filling. Sedimentology 47, 121–178. doi:10.1046/j.1365-3091.2000.00006.x

Park, H., Scheidt, C., Fenwick, D., Boucher, A., Caers, J., 2013. History matching and uncertainty quantification of facies models with multiple geological interpretations. Comput. Geosci. 17, 609–621. doi:10.1007/s10596-013-9343-5

Peredo, O., Ortiz, J.M., Herrero, J.R., Samaniego, C., 2014. Tuning and hybrid parallelization of a genetic-based multi-point statistics simulation code. Parallel Comput. doi:10.1016/j.parco.2014.04.005

Perulero Serrano, R., Guadagnini, L., Riva, M., Giudici, M., Guadagnini, A., 2014. Impact of two geostatistical hydro-facies simulation strategies on head statistics under non-uniform groundwater flow. J. Hydrol. 508, 343–355. doi:10.1016/j.jhydrol.2013.11.009

Pirot, G., Straubhaar, J., Renard, P., 2014. Simulation of braided river elevation model time series with multiple-point statistics. Geomorphology 214, 148–156. doi:10.1016/j.geomorph.2014.01.022

Pyrcz, M.J., Boisvert, J.B., Deutsch, C.V., 2008. A library of training images for fluvial and deepwater reservoirs and associated code. Comput. Geosci. 34, 542–560. doi:10.1016/j.cageo.2007.05.015

Pyrcz, M.J., Boisvert, J.B., Deutsch, C.V., 2009. ALLUVSIM: A program for event-based stochastic modeling of fluvial depositional systems. Comput. Geosci. 35, 1671–1685. doi:10.1016/j.cageo.2008.09.012

Ramanathan, R., Guin, A., Ritzi, R.W., Dominic, D.F., Freedman, V.L., Scheibe, T.D., Lunt, I.A., 2010. Simulating the heterogeneity in braided channel belt deposits: 1. A geometric-based methodology and code. Water Resour. Res. 46, W04515. doi:10.1029/2009WR008111

RamaRao, B.S., Lavenue, M., de Marsily, G., Marietta, M.G., 1995. Pilot point methodology for automated calibration of an ensemble of conditionally simulated transmissivity fields: 1. Theory and computational experiments. Water Resour. Res. 31, 475–493.

Refsgaard, J.C., Auken, E., Bamberg, C.A., Christensen, B.S.B., Clausen, T., Dalgaard, E., Effersø, F., Ernstsen, V., Gertz, F., Hansen, A.L., He, X., Jacobsen, B.H., Jensen, K.H., Jørgensen, F., Jørgensen, L.F., Koch, J., Nilsson, B., Petersen, C., De Schepper, G., Schamper, C., Sørensen, K.I., Therrien, R., Thirup, C., Viezzoli, A., 2014. Nitrate reduction in geologically heterogeneous catchments — A framework for assessing the scale of predictive capability of hydrological models. Sci. Total Environ. 468–469, 1278–1288. doi:10.1016/j.scitotenv.2013.07.042

Rimstad, K., and H. Omre, 2010, Impact of rock-physics depth trends and Markov random fields on hierarchical Bayesian lithology/ fluid prediction: Geophysics, 75, no. 4, R93–R108

Rongier, G., Collon-Drouaillet, P., Filipponi, M., 2014. Simulation of 3D karst conduits with an object-distance based method integrating geological knowledge. Geomorphology. doi:10.1016/j.geomorph.2014.04.024

Rubin, Y., Mavko, G., Harris, J., 1992. Mapping permeability in heterogeneous aquifers using hydrologic and seismic data, Water Resour. Res., 28(7), 1809–1816, doi:10.1029/92WR00154.

Sahni, I., Horne, R.N., 2005. Multiresolution wavelet analysis for improved reservoir description. SPE Reserv. Eval. Eng. 8, 53–69. doi:10.2118/87820-PA

Sambridge, M., 1999a. Geophysical inversion with a neighbourhood algorithm—I. Searching a parameter space. Geophys. J. Int. 138, 479–494. doi:10.1046/j.1365-246X.1999.00876.x

Sambridge, M., 1999b. Geophysical inversion with a neighbourhood algorithm—II. Appraising the ensemble. Geophys. J. Int. 138, 727–746. doi:10.1046/j.1365-246x.1999.00900.x

Sambridge, M., 2014, A parallel tempering algorithm for probabilistic sampling and multimodel optimization. Geophys. J. Int. 196, 357-374. doi:10.1093/gji/ggt342.

Sarma, P., Chen, W. H., 2009. Generalization of the ensemble kalman filter using kernels for non-Gaussian random fields. SPE Reservoir Simulation Symposium, 1146-1165.

Sarma, P., Durlofsky, L.J., Aziz, K., 2008. Kernel Principal Component Analysis for efficient, differentiable parameterization of multipoint geostatistics. Math. Geosci. 40, 3–32. doi:10.1007/s11004-007-9131-7

Scheibe, T.D., Chien, Y.-J., 2003. An evaluation of conditioning data for solute transport prediction. Ground Water 41, 128–141. doi:10.1111/j.1745-6584.2003.tb02577.x

Scheibe, T.D., Freyberg, D.L., 1995. Use of sedimentological information for geometric simulation of natural porous media structure. Water Resour. Res. 31, 3259–3270.

Scheidt, C., Jeong, C., Mukerji, T., and Caers, J., 2015, Probabilistic falsification of prior geologic uncertainty with seismic amplitude data: Application to a turbidite reservoir case, Geophysics, 80, M89-M12.

Schoups, G., Vrugt, J.A., 2010. A formal likelihood function for parameter and predictive inference of hydrologic models with correlated, heteroscedastic, and non-Gaussian errors. Water Resour. Res. 46, W10531. doi:10.1029/2009WR008933

Sen, M.K., Stoffa, P.L., 2013. Global Optimization Methods in Geophysical Inversion. Cambridge University Press.

Stein, M., Kolbjornsen, O., 2011. Facies modeling using a Markov mesh model specification. Math. Geosci., 43, 611-624.

Straubhaar, J., Renard, P., Mariethoz, G., Froidevaux, R., Besson, O., 2011. An improved parallel multiple-point algorithm using a list approach. Math. Geosci. 43, 305–328. doi:10.1007/s11004-011-9328-7

Straubhaar, J., Walgenwitz, A., Renard, P., 2013. Parallel multiple-point statistics algorithm based on list and tree structures. Math. Geosci. 45, 131–147. doi:10.1007/s11004-012-9437-y

Strebelle, 2002. Conditional simulation of complex geological structures using multiple-point statistics. Math. Geol. 34, 1–21. doi:10.1023/A:1014009426274

Strebelle, S., Cavelius, C., 2014. Solving speed and memory issues in multiple-point statistics simulation program SNESIM. Math. Geosci. 46, 171–186. doi:10.1007/s11004-013-9489-7

Suzuki, S., Caers, J., 2008. A distance-based prior model parameterization for constraining solutions of spatial inverse problems. Math. Geosci. 40, 445–469. doi:10.1007/s11004-008-9154-8

Tahmasebi, P., Hezarkhani, A., Sahimi, M., 2012a. Multiple-point geostatistical modeling based on the cross-correlation functions. Comput. Geosci. 16, 779–797. doi:10.1007/s10596-012-9287-1

Tahmasebi, P., Sahimi, M., Mariethoz, G., Hezarkhani, A., 2012b. Accelerating geostatistical simulations using graphics processing units (GPU). Comput. Geosci. 46, 51–59. doi:10.1016/j.cageo.2012.03.028

Tarantola, A., 2005. Inverse problem theory and methods for model parameter estimation. Society for Industrial and Applied Mathematics, Philadelphia, PA.

Tarantola, A., 2006. Popper, Bayes and the inverse problem. Nat. Phys. 2, 492–494.

Tarantola, A., Valette, B., 1982. Inverse problems - Quest for information. J. Geophys. 159–170.

Tikhonov, A.N., Arsenin, V.I., 1977. Solutions of ill-posed problems. Winston.

Ulvmoen, M., and H. Omre, 2010, Improved resolution in Bayesian lithology/fluid inversion from prestack seismic data and well observations: Part 1 — Methodology  Geophysics 75, R21-R35.

Vrugt, J.A., ter Braak, C.J.F., Diks, C.G.H, Robinson, B.A., Hyman, J.M., Higdon, D., 2009, Accelerating Markov chain Monte Carlo simulation by differential evolution with self-adaptive randomized subspace sampling. International Journal of Nonlinear Sciences and Numerical Simulation, 10, 273-290.

Webb, E.K., Anderson, M.P., 1996. Simulation of preferential flow in three-dimensional, heterogeneous conductivity fields with realistic internal architecture. Water Resour. Res. 32, 533–545.

Wellmann, J.F., Finsterle, S., Croucher, A., 2014. Integrating structural geological data into the inverse modelling framework of iTOUGH2. Comput. Geosci. 65, 95–109. doi:10.1016/j.cageo.2013.10.014

Zhang, T., Switzer, P., Journel, A., 2006. Filter-based classification of training image patterns for spatial simulation. Math. Geol. 38, 63–80. doi:10.1007/s11004-005-9004-x

Zheng, C., Gorelick, S.M., 2003. Analysis of solute transport in flow fields influenced by preferential flowpaths at the decimeter scale. Ground Water 41, 142–155. doi:10.1111/j.1745-6584.2003.tb02578.x

Zhou, H., Gómez-Hernández, J., Li, L. 2012. A pattern search based inverse method. Water Resources Research, 48, W03505.

Zhou, H., Gómez-Hernández, J. J., Hendricks Franssen, H. J. & LI, L. 2011. An approach to handling non-Gaussianity of parameters and state variables in ensemble Kalman filtering. Advances in Water Resources, 34, 844-864.

Zinn, B., Harvey, C.F., 2003. When good statistical models of aquifer heterogeneity go bad: A comparison of flow, dispersion, and mass transfer in connected and multivariate Gaussian hydraulic conductivity fields. Water Resour. Res. 39, 1051. doi:10.1029/2001WR001146